\documentclass[11pt,aps,prd,nofootinbib]{revtex4-1}
\usepackage{latexsym,bm,amsmath,amssymb,graphicx,subfigure,tensind}
\usepackage{tabularx}
\usepackage{titlesec}
\usepackage[linktoc=all,hidelinks]{hyperref}
\titleformat{\section}
{\normalfont\large\bfseries}{\thesection}{1em}{}
\titleformat{\subsection}
{\normalfont\normalsize\bfseries}{\thesubsection}{1em}{}
\titleformat{\subsubsection}
{\normalfont\normalsize\bfseries}{\thesubsubsection}{1em}{}

\numberwithin{equation}{section}

\tensordelimiter{?}
\tensorformat{}

\numberwithin{equation}{section}
\usepackage{hyperref}
 \hypersetup{
     colorlinks=true,
     linkcolor=black,
     filecolor=blue,
     citecolor=red,      
     urlcolor=blue,
     }
\usepackage[most]{tcolorbox}
\newcommand{\link}[1]{[\href{http://arxiv.org/abs/#1}{{\tt arXiv:#1}}]}
\newcommand{\linkth}[1]{[\href{http://arxiv.org/abs/hep-th/#1}{{\tt arXiv/hep-th:#1}}]}
\newcommand{\linkgr}[1]{[\href{http://arxiv.org/abs/gr-qc/#1}{{\tt arXiv/gr-qc:#1}}]}
\renewcommand{\thesection}{\arabic{section}}
\renewcommand{\thesubsection}{\arabic{section}.\arabic{subsection}}
\renewcommand{\thesubsubsection}{\arabic{section}.\arabic{subsection}.\arabic{subsubsection}}
\DeclareMathAlphabet{\mathpzc}{OT1}{pzc}{m}{it}

\newcommand{\ah}{\hat{a}}
\newcommand{\bh}{\hat{b}}
\newcommand{\ch}{\hat{c}}
\newcommand{\hh}{\hat{h}}
\newcommand{\vh}{\hat{v}}
\newcommand{\nb}{\bar{\nabla}}
\newcommand{\Ab}{\bar{A}}
\newcommand{\Fb}{\bar{F}}

\newcommand{\eh}{\hat{e}}

\newcommand{\fc}{\mathfrak{c}}
\newcommand{\ft}{\mathfrak{t}}
\newcommand{\fb}{\mathfrak{b}}

\newcommand{\fs}{\mathfrak{s}}

\begin{document}

\title{Holographic stress tensor of colored Lifshitz spacetimes and hairy black holes}
\author{Deniz O. Devecio\u{g}lu}
\email{dodeve@gmail.com}
\affiliation{School of Physics, Huazhong University of Science and Technology,\\
             Wuhan, Hubei,  430074, China}

\date{\today}

\begin{abstract}
We compute the holographic stress tensor of colored Lifshitz spacetimes following the proposal by Ross-Saremi for gravity duals of non-relativistic theories. For a well-defined variational principle, we first construct a finite on-shell action for the Einstein-Yang-Mills model in four dimensions with Lifshitz spacetime as a solution. We then solve the linearised equations of motion and identify the modes that preserve the asymptotically Lifshitz condition. Employing these modes, we also show that the stress tensor is finite, obeying the scaling and the diffeomorphism Ward identities, i.e., conservations laws. As a final application, we evaluate the energy density and the spatial stress tensor of the previously found numerical black hole solutions with various dynamical exponents $z$. The alternative Smarr relation that has been used in Lifshitz black holes and the first law of thermodynamics are shown to hold without a global Yang-Mills charge, indicating the black holes in question are hairy.
\end{abstract}
 \maketitle
 \tableofcontents
\newpage
\section{Introduction}\label{intro}
A great deal of non-relativistic systems exhibit dynamical scaling at the fixed points of their renormalisation group flow
\begin{align}
t\rightarrow \lambda^{z}t,\quad x^{i}\rightarrow\lambda x^{i},\quad r\rightarrow \lambda^{-1}r, \quad z>1,
\end{align}
which can be seen as a generalisation of the conformal scaling experienced in relativistic theories. Following the logic of gauge/gravity duality, one can then look for spacetimes that are weakly coupled dual descriptions of the strongly coupled field theories enjoying this anisotropic scale invariance \cite{Son:2008ye, Balasubramanian:2008dm, Kachru:2008yh}. In addition to the scale symmetry, demanding invariance under space and time translations, spatial rotations will lead to the Lifshitz spacetimes in the bulk \cite{Kachru:2008yh}
\begin{align}
 ds^{2}=-r^{2z} dt^{2}+r^{2}(d x^{2}+d y^{2})+\dfrac{dr^{2}}{r^{2}}\label{lifshitz},
\end{align}
which has curios geometric properties. For example, unlike AdS spacetimes, where there is a well-defined boundary in the form of a conformal structure, the Lifshitz spacetimes do not have a non-degenerate metric at the boundary \cite{Horava:2010vho, Copsey:2010ya}. Therefore, the usual tools of holography like the Fefferman-Graham (FG) expansion do not work, and others need a modification to work in the non-relativistic case. There has been a flurry of activity on developing notions and tools for non-relativistic holography (see the review \cite{Taylor:2015glc} and references therein e.g., \cite{Chemissany:2014xsa,Mann:2011hg,Ross:2011gu,Ross:2009ar,Christensen:2013rfa,Christensen:2013lma,Hartong:2014oma}). In this regard, the non-relativistic stress-energy tensor \cite{Ross:2009ar}, inspired by \cite{Hollands:2005ya} will be our primary tool in this work. To put it briefly, instead of working with the boundary metric, the stress tensor will be defined by varying the boundary frame fields and keeping the tangent space components of the of matter fields fixed. While being more suited to a non-relativistic setting, this approach is also crucial for defining correct conserved charges for gravity theories with matter fields other than scalars. In Section \ref{stresssec} we will discuss the details of this construction.

Another challenging feature of these spacetimes is that the cosmological Einstein theory does not admit these as a solution; therefore, some type of matter coupling or higher curvature terms should be considered \cite{Taylor:2008tg, Cai:2009ac, Ayon-Beato:2009rgu, Brenna:2011gp, Lu:2012xu}, making the holographic analysis more intricate. An examplar is the massive vector model \cite{Taylor:2008tg}, which is simpler and includes all $z>1$ as a solution. This system was the testing ground for the non-relativistic stress tensor, where it was shown to generate finite results for linearised perturbations \cite{Ross:2009ar}. Later, building on this work, a definition of asymptotically locally Lifshitz spacetimes is given, and the holographic renormalisation of one-point functions is studied in \cite{Ross:2011gu}.  Another important model is the $z=2$ specific, four-dimensional system which is obtained from the Scherk-Schwarz reduction of a five-dimensional axion-dilaton gravity living on an asymptotically locally AdS spacetime \cite{Christensen:2013rfa, Christensen:2013lma}. By computing the well-established higher dimensional AdS sources+VEVs, Ward identities, the corresponding $z=2$ Lifshitz holographic dictionary was identified. A major result of this work was the appearance of torsional Newton-Cartan (TNC) geometry at the boundary \cite{Hartong:2015wxa, Hartong:2014oma}.

In an attempt to further extend the Lifshitz holography, understand different matter couplings better and search for dynamical curiosities, the main focus of this work will be the four-dimensional Einstein-Yang-Mills (EYM) model. The $SU(2)$ colored Lifshitz background solutions was first found in \cite{Devecioglu:2014iia}, later extended to five dimensions in \cite{Fan:2015yza}. In fact, the embedding of four-dimensional EYM in eleven-dimensional supergravity was shown quite a long time ago \cite{Pope:1985bu}, where the dynamical exponent for the corresponding Lifshitz solutions should be fixed to an irrational number $z=1+\sqrt{6}$ as a supersymmetry requirement \cite{Fan:2015yza}.

Our first step towards a holographic stress tensor for the EYM theory will be to find the boundary terms to have a well-defined variational problem with an extremized action on four-dimensional Lifshitz background solutions. This is performed in Section \ref{boundarycounter} by considering the Hawking-Gibbons (HG) term and a counter-term intrinsic to the boundary. Next, in order to find the perturbative modes that contribute to the stress tensor in the dual field theory, we solve the linearised field equations and identify the modes that lead to the Lifshitz background \eqref{lifshitz}. The identification of modes will be an important stepping stone for further analysis, e.g., the holographic renormalisation of the one-point functions. Through perturbation solutions, we show that the stress tensor stays finite and obeys the Ward identities. This is done for both constant and general perturbations in Section \ref{PERT}. For constant perturbations, it is only possible to show that the scaling Ward identity holds, as there is no dependence on time or boundary coordinates. On the other hand, with general perturbations, we will show that the rest of the conservation laws are satisfied by incorporating an expansion in wavenumber $k$ and frequency $\omega$. 

Leaving aside the possible applications in non-relativistic holography, the EYM solutions (both particle-like and black holes) are certainly interesting in their own rights. The asymptotically flat solutions by Bartnik-McKinnon \cite{Bartnik:1988am} was quite a surprise for the community since Deser's proof of no static solutions to the Yang-Mills (YM) equations in four dimensions \cite{Deser:1976wq}, and Lichnerowicz theorem of no gravitational solitons \cite{Lichner}  suggest that the EYM system do not admit particle-like solutions. However, the non-linearities of both the gravity and the gauge system together make the solutions possible. Besides particle-like solutions, the EYM model also admits colored black holes in asymptotically flat \cite{Bizon:1990sr,Volkov:1998cc} and AdS cases \cite{Torii:1995wv,Winstanley:1998sn}, which are hairy and violates the no-hair theorem. 

On the Lifshitz side, the colored numerical black holes, which are parametrised by the fine-tuned gauge field strength at the horizon, was constructed for different horizon topologies \cite{Devecioglu:2014iia}. The behaviour of solutions differs for black holes that are large or small with respect to the length scale $L$, which is fixed by the cosmological constant. In some ways, the solutions are quite different from their conformal and asymptotically flat cousins, e.g., for a given horizon radius $R_{0}$ and dynamical exponent $z$, there is a unique value of the gauge field that leads to desired Lifshitz asymptotics. This is in contrast to the solutions with AdS asymptotics, where the solutions exist for a value of shooting parameter in continuous open intervals \cite{Winstanley:2008ac}. Apart from their functional behaviour, the physics of these black holes are not studied. Therefore, the second part of this work is devoted to the conserved quantities and the thermodynamics of these black holes. 

The generalised Smarr relation \cite{Kastor:2009wy} for Lifshitz black holes that is valid for all horizon topologies reads
\begin{align}
(D-3)M=(D-2)TS-2PV, \label{origsmarr}
\end{align}
along with the first law
\begin{align}
dM=TdS+VdP,
\end{align}
where $P,\,V$ are thermodynamic pressure and volume, respectively. As we pointed out before, the Lifshitz asymptotics require matter, which in turn makes the cosmological constant dependent on the couplings. This obscures the definition of thermodynamic pressure, so instead, some of the works used alternative Smarr relation paired with the first law as follows \cite{Liu:2014dva,Bertoldi:2009dt,Dehghani:2010kd,Berglund:2011cp,Dehghani:2011hf,Dehghani:2013mba}
\begin{align}
(D+z-2)M=(D-2)TS,\quad dM=TdS.\label{altsmarr}
\end{align} 
Later, the alternative relation \eqref{altsmarr} was shown to be consistent with the original one \eqref{origsmarr} when the horizon topology is planar \cite{Brenna:2015pqa}. Thus, we will make use of the reduced relation in the study of the thermodynamics of the planar numerical solutions. One of the important results in our study of thermodynamics will be the hairy nature of the solutions. Actually, the lack of global YM charge was argued by linear analysis of field equations \cite{Fan:2015yza}. In Section  \ref{numblackhole} we will support this result by showing the first law of thermodynamics and the Smarr relation \eqref{altsmarr} hold without any global YM charge.  

The outline of the paper is as follows: In Section \ref{boundarycounter} we set up our model and find out the boundary terms for a well-defined variational principle. Then in Section \ref{stresssec}, we review the holographic stress-energy tensor for non-relativistic theories following the recipe by Ross-Saremi \cite{Ross:2009ar}. We then set the stage for perturbation analysis by choosing a suitable basis for perturbation of the frame fields and the gauge field components in Section \ref{pertsetup}. The following two Sections \ref{constantperturbations}, \ref{generalpert} will be reserved to the solutions and the stress tensor computations of constant perturbations and general perturbations, respectively. Section \ref{numblackhole} deals with the energy and thermodynamics of the numerical black hole solutions of \cite{Devecioglu:2014iia}. Finally we conclude in Section \ref{conclusions}. Explicit results for the linearisation of some expressions and the field equations of generalised perturbations are presented in Appendix \ref{appendix1}-\ref{appendix2}.

Since the number and the type of indices are somewhat involved, let us carefully display our conventions (along the way there will be several addendeums). Here $ F_{\mu\nu}$ is the gauge field strength $F_{\mu\nu}\equiv F_{\mu\nu}^{\Lambda}\,T_{\Lambda}=\partial_{\mu}A_{\nu}-\partial_{\nu}A_{\mu}-i[A_{\mu},A_{\nu}]$ and we choose generators $T_{\Lambda}\equiv \tau_{\Lambda}/2,\,\, \Lambda=1,2,3$ with $\tau_{\Lambda}$ denoting Pauli matrices with upper case Greek indices for the Yang-Mills algebra. The commutation relations and the normalization of generators are
given as $ [T_{\Lambda},T_{\Gamma}]=i\epsilon_{\Lambda\Gamma\Delta}T_{\Delta}$ and $ \text{Tr}(\,T_{\Lambda}T_{\Gamma})=\delta_{\Lambda\Gamma}/2$, respectively. For the the signature of metric we use $(-,+,+,+)$, and the Riemann tensor is taken as
$R^{\mu}\,_{\nu\alpha\beta}=\partial_{\alpha} \Gamma^{\mu}\,_{\beta\nu}-\cdots$ with
$R_{\mu\nu}=R^{\alpha}\,_{\mu\alpha\nu}$. We also use $x^{\mu}=(t,x_{i},r)$ for spacetime (curved) indices and $y^{a}=(t,x_{i})$ for boundary coordinate indices. Finally we use  $i,j=1,2$ indices for the planar part. The integration constants of solutions will be displayed in fraktur font e.g. $\fb_{1},\fc_{2},\fs_{3},\cdots$.
\section{Boundary counterterms in EYM }{\label{boundarycounter}}
We start with the four-dimensional Einstein gravity with a cosmological constant minimally coupled to $SU(2)$ gauge fields  described by the action
\begin{equation}
 S_{\mathcal{M}}=\int_\mathcal{M} d^{4}x\,\sqrt{-g}\Big((R-2\Lambda)-\dfrac{1}{2g_{\text{\tiny YM}}^{2}} \text{Tr}\left[\,F_{\mu\nu}F^{\mu\nu}\right]\Big)\label{action},\,
\end{equation}
where $\Lambda$ is the cosmological and $g_{\text{\tiny YM}}^{2}$ is the gauge coupling constant in dimensions of
$1/\text{length}^{2}$. Unlike EYM solutions in AdS, where the background is already a solution to the cosmological Einstein theory; here, we use matter to support backgrounds with anisotropic scaling symmetry.

The variation of (\ref{action}) will amount to
\begin{align}
\delta S_{\mathcal{M}}= \int_{\mathcal{M}} d^{4}x\,\sqrt{-g} (\mathcal{E}_{\mu\nu}\,\delta g^{\mu\nu}+\mathcal{E}^{\Lambda}_{\mu}\,\delta A_{\Lambda}^{\mu})+ \text{boundary terms},\label{varaction}
\end{align}
where $\mathcal{E}_{\mu\nu}$ is the field equations for the metric
\begin{align}
\mathcal{E}_{\mu\nu}\equiv R_{\mu\nu}-\frac{1}{2}R g_{\mu\nu}+\Lambda g_{\mu\nu}
-\frac{1}{g_{\text{\tiny YM}}^{2}}T_{\mu\nu}\label{Geqns},
\end{align}
with the traceless YM stress energy tensor defined as
\begin{align}
 T_{\mu\nu}\equiv\text{Tr}\, (F_{\mu}\,^{\alpha}F_{\nu\alpha}-\dfrac{1}{4}g_{\mu\nu}F_{\alpha\beta}F^{\alpha\beta}).\label{stress}
\end{align}
For the matter sector, the field equations are
\begin{align}
 \mathcal{E}_{\mu}^{\Lambda}\equiv D^{\nu}F_{\nu\mu}^{\Lambda}=0,\label{ymeqns}
\end{align}
where the gauge covariant derivative is $D_{\mu}\equiv \nabla_{\mu}-i[A_{\mu},\quad]$.

We assume a Lifshitz background solution of the form 
\begin{align}
 ds^{2}=-r^{2z} dt^{2}+r^{2}(d x^{2}_{1}+d x^{2}_{2})+\dfrac{dr^{2}}{r^{2}}\label{backmet},
\end{align}
paired with the planar symmetric $SU(2)$ gauge field ansatz \cite{Basler:1986yr}
\begin{align}
A_{\mu}d x^{\mu}= Q(r) T^{3} dt+ R(r) T^{1} dx_{1}+  R(r) T^{2} dx_{2}.
\label{planarsymgauge}
\end{align}
Since we look for a background solution, we assume all functions depend on the radial coordinate $r$, and for simplicity, we focus on the purely magnetic case, i.e. $Q(r)=0$. Then through field equations (\ref{Geqns}), (\ref{ymeqns}), the gauge field solution is found to be
\begin{align}
 R(r)=\pm \sigma r\,,\quad \text{for all}\,\, z > 1,\quad \sigma\equiv\sqrt{z+1}\label{backsol},
\end{align}
provided the cosmological constant and the gauge coupling are chosen as follows
\begin{align}
\Lambda=-\dfrac{3+2z+z^2}{2},\quad g_{\text{\tiny YM}}^{2}=\frac{1}{2}\dfrac{(z+1)}{(z-1)}\label{const}.
\end{align}
The generalization of this solution to five dimensions and the exact black hole solutions with an extra Maxwell charge is found in \cite{Fan:2015yza}. In the conformal limit $z\rightarrow 1$, the YM part decouples from the gravity action, and the decoupled gauge field is a solution to the pure YM. The sign ambiguity of the gauge field corresponds to a gauge transformation. Hence in what follows, we will continue with the positive sign gauge field.

Having reviewed the Lifshitz solution, let us construct the appropriate action which satisfies $\delta S=0$ on-shell. The boundary terms in (\ref{varaction}) consist of the well known gravitational part $g^{\mu\nu}\delta R_{\mu\nu}=\nabla_{\mu}f^{\mu}$ and the matter part
\begin{align}
\delta S_{\mathcal{M}}=\text{field equations}+\int_{\mathcal{M}}  d^{4}x\,\sqrt{-g} \nabla^{\mu}\left( f_{\mu}-\dfrac{1}{g_{\text{\tiny YM}}^{2}} F^{\Lambda}_{\mu\nu}\delta A_{\Lambda}^{\nu} \right),\label{varSM}
\end{align}
where
\begin{align}
f^{\mu}\equiv g^{\rho\nu}\delta\Gamma^{\mu}_{\rho\nu}-g^{\mu\nu}\delta\Gamma^{\rho}_{\rho\nu}.
\end{align}
To cancel out the metric variations with derivatives and have a boundary value problem with Dirichlet conditions at $\partial\mathcal{M}$, the addition of the trace of extrinsic curvature known as the Hawking-Gibbons (HG) term is necessary. However, this alone is not sufficient to pose a well-defined variational problem in holography. We also require boundary counterterms to extremize the action on solutions of the equations of motion. Taking these and the gauge invariance into account, one is naturally led to choose
\begin{align}
S=S_{\mathcal{M}}+S_{\partial{\mathcal{M}}}&= \int_{\mathcal{M}} d^{4}x\,\sqrt{-g}\Big((R-2\Lambda)-\dfrac{1}{2 g_{\text{\tiny YM}}^{2}} \text{Tr}[F_{\mu\nu}F^{ \mu\nu}]\Big)\nonumber\\
&+\int_{\partial{\mathcal{M}}} d^{3}y \,\sqrt{-\gamma}\left(2 K +k_{0}+ \alpha\, \text{Tr}\left[F_{ab}F^{ab}\right]\right)\label{action3},\,
\end{align}
where $y^{a}$ denotes the coordinates on the boundary at some constant $r$, $\gamma_{ab}$ is the induced metric, $k_{0}$ and $\alpha$ are the constants to be determined. The extrinsic curvature tensor is given as $K_{ab}=\nabla_{(a}n_{b)}$  where $n^{\mu}=(0,0,0,r)$ is an outward-directed unit vector orthogonal to the boundary. The antisymmetry of $F_{\mu\nu}$ makes it possible to define the following tensor with a boundary index
\begin{align}
n_{\mu}F^{\mu\nu}_{\Lambda} \rightarrow n_{\mu}F^{\mu a}_{\Lambda}\equiv E^{a}_{\Lambda},
\end{align}
which we use to write down (\ref{varSM}) as
\begin{align}
\delta S_{\mathcal{M}}= \text{field equations}
+ \int_{\partial\mathcal{M}}  d^{3}y\,\sqrt{-\gamma} \left( n_{\mu}f^{\mu}-\dfrac{1}{g_{\text{\tiny YM}}^{2}} E_{\Lambda}^{a}\delta A^{\Lambda}_{a} \right).\label{varbulk1}
\end{align}
The boundary contributions from the HG part of $S_{\partial{\mathcal{M}}}$ is
\begin{align}
\delta\left[\int_{\partial{\mathcal{M}}} d^{3}y \,\sqrt{-\gamma}\left(2 K +k_{0}\right)\right]=&\int_{\partial{\mathcal{M}}} d^{3}y \,\sqrt{-\gamma}\left(-f^{\mu}n_{\mu}+\left[\Pi_{ab}-\frac{k_{0}}{2}\gamma_{ab}\right]\delta \gamma^{ab}\right)\label{Kvarpiece}
\end{align}
where we defined $\Pi_{ab}\equiv K_{ab}-K \gamma_{ab}$. The first term in \eqref{Kvarpiece} handles the variations with derivatives and for the rest, we need the components of the extrinsic curvature of the solution (\ref{backmet})
\begin{align}
K_{tt}=-zr^{2z},\quad K_{ij}=r^{2}\delta_{ij},\quad K=z+2.
\end{align}
Likewise, the matter part of the variation reads 
\begin{multline}
\delta\left(\int_{\partial{\mathcal{M}}} d^{3}y \,\sqrt{-\gamma}\,\alpha\,\text{Tr}\left[F_{ab}F^{ab}\right]\right)=\alpha\int_{\partial{\mathcal{M}}} d^{3}y \,\sqrt{-\gamma}\Big\{ (F_{ac}^{\Lambda}F_{b}{}^{c}_{\Lambda}-\frac{1}{4} F_{cd}^{\Lambda}F_{\Lambda}^{cd}\gamma_{ab})\delta \gamma^{ab}\\-2 \epsilon^{\Lambda\Gamma}{}_{\Delta}A_{\Gamma}^{b}F_{ba}^{\Delta}\delta A^{a}_{\Lambda}\Big\}.\label{betaterms}
\end{multline}
We also define the following general variation for future convenience
\begin{align}
\delta S= \int d^{3}y (s_{ab}\delta \gamma^{ab}+s^{\Lambda}_{a}\delta A^{a}_{\Lambda}),\label{genvar}
\end{align}
where $s_{ab}$ and $s_{a}$ are 
\begin{align}
s_{ab}\equiv &\sqrt{-\gamma}\left[\Pi_{ab}-\frac{k_{0}}{2}\gamma_{ab}+\alpha\left(F_{ac}^{\Lambda}F_{b}{}^{c}_{\Lambda}-\frac{1}{4} F_{cd}^{\Lambda}F_{\Lambda}^{cd}\gamma_{ab}\right)\right],\\
s^{\Lambda}_{a}\equiv & -\sqrt{-\gamma}\left[\frac{1}{g_{\text{\tiny YM}}^{2}}E^{\Lambda}_{a}+2\alpha \epsilon^{\Lambda\Gamma}{}_{\Delta}A^{b}_{\Gamma} F^{\Delta}_{ba}\right].
\end{align}
Here the constants $k_{0}$ and $\alpha$ will be chosen such that $s_{ab}$ and $s_{a}^{\Lambda}$ will be equal to zero on solutions making $\delta S=0$ for arbitrary variations around \eqref{backmet}, \eqref{backsol}. With this in mind, evaluating the solution on individual boundary terms we have
\begin{align}
\dfrac{1}{g_{\text{\tiny YM}}^{2}} E_{a}^{\Lambda}\,\delta A_{\Lambda}^{a}&=\dfrac{1}{g_{\text{\tiny YM}}^{2}}r\,\sqrt{z+1}(\delta A_{1}^{x_{1}}+\delta A_{2}^{x_{2}}),\\
\left[\Pi_{ab}-\frac{k_{0}}{2}\gamma_{ab}\right]\delta \gamma^{ab}&=\frac{1}{2}r^{2z}(4+k_{0})\delta \gamma^{tt}-\frac{1}{2}r^{2}(2(z+1)+k_{0})\delta_{ij}\delta \gamma^{ij},\\
\alpha(F_{ac}^{\Lambda}F_{b}{}^{c}_{\Lambda}-\frac{1}{4} F_{cd}^{\Lambda}F_{\Lambda}^{cd}\gamma_{ab})\delta \gamma^{ab}&=\frac{\alpha}{2}r^{2z}(z+1)^{2}\delta \gamma^{tt}+\frac{\alpha}{2}r^{2}(z+1)^{2}\delta_{ij}\delta \gamma^{ij},\\
-2\alpha \epsilon^{\Lambda\Gamma}{}_{\Delta}A^{b}_{\Gamma} F^{\Delta}_{ba}\delta A_{\Lambda}^{a}&=-2 \alpha r\,(z+1)^{3/2}(\delta A_{1}^{x_{1}}+\delta A_{2}^{x_{2}}).
\end{align}
Gathering all these pieces the general variation (\ref{genvar}) reads
\begin{multline}
\delta S=\int_{\partial\mathcal{M}}  d^{3}y\,\sqrt{-\gamma} \Bigg\{\frac{1}{2}r^{2z} \left(4+k_{0}+\alpha (z+1)^{2}\right)\delta \gamma^{tt}+\frac{1}{2}r^{2}\Big((z+1)(\alpha(z+1)-2)-k_{0}\Big)\delta^{ij}\delta \gamma_{ij}\nonumber\\
+\frac{r\sqrt{z+1}}{g_{\text{\tiny YM}}^{2}}\left[2\alpha g_{\text{\tiny YM}}^{2}(z+1)-1\right](\delta A^{1}_{x_{1}}+\delta A^{2}_{x_{2}})\Bigg\}.
\end{multline}
Provided $k_{0}=-(z+3)$ and $\alpha=(z-1)/(z+1)^{2}$, the above expression is zero and we have a well defined variational problem. Therefore our final action is the following
\begin{align}
S=&\int_{\mathcal{M}}d^{4}x\sqrt{-g}\left((R-2\Lambda)-\frac{\text{Tr}[F_{\mu\nu}F^{ \mu\nu}]}{2g_{\text{\tiny YM}}^{2}}\right)\nonumber\\
+&\int_{\partial\mathcal{M}}d^{3}y\sqrt{-\gamma}\left(2K-(z+3)+\frac{\text{Tr}[F_{ab}F^{ab}]}{2g_{\text{\tiny YM}}^{2}(z+1)}\right).
\end{align}

Note that in addition to the boundary terms we have considered, one can include terms involving derivatives of the boundary fields, which we denote as $S_{deriv}$. However, the scalars constructed from the boundary fields are constant for the Lifshitz solution, and with a planar boundary metric, the derivatives of these scalars do not contribute to the background (\ref{backmet}). In the case of general asymptotically Lifshitz spacetimes, the derivative terms can be considered rendering the components of the stress tensor finite if needed \cite{Ross:2009ar}. Furthermore, inspired by the Einstein-Maxwell model, one can be tempted to consider the following boundary term for the matter part \cite{Braden:1990hw}
\begin{align}
\int_{\partial{\mathcal{M}}} d^{3}y \,\sqrt{-\gamma}\,\text{Tr}\left[n^{\mu}\,F_{\mu\nu}A^{ \nu}\right].
\end{align}
However, unlike its abelian counterpart, this choice is not gauge invariant on-shell, so our choice seems to be the only plausible one.

Having determined the boundary counterterms for EYM action with Lifshitz asymptotics, we can now review the construction of stress tensor introduced in \cite{Ross:2009ar} for non-relativistic spacetimes. 

\section{Stress tensor for non-relativistic theories}{\label{stresssec}}
This section will recapitulate the procedure given in \cite{Ross:2009ar} for calculating a boundary stress tensor from a bulk action principle in non-relativistic theories. The construction is analogous to the relativistic case \cite{Hollands:2005ya} and relies on the definition of a modified stress tensor in the presence of non-scalar boundary fields. We will first work at a finite cut-off $r_{c}$, then take $r_{c}\rightarrow\infty$ for the final result, as there is no non-degenerate metric at the boundary.

The following definition is sufficient when the metric is the only non-trivial boundary field in relativistic theory \cite{Balasubramanian:1999re} 
\begin{align}
T^{ab}\hat{\epsilon}=-2\frac{\delta S}{\delta \gamma_{ab}}\label{emten1},
\end{align}
where $\hat{\epsilon}$ is the volume form associated with $\gamma_{ab}$. The covariant conservation of (\ref{emten1}) can be shown by considering the action of boundary diffeomorphisms on the variation. This definition was modified by Hollands-Ishibashi-Marolf when the theory in question involves extra fields other than scalars. Following the recipe in \cite{Hollands:2005ya}, it is convenient to consider the set of frame fields at the boundary
\begin{align}
\gamma_{ab}=\eh^{M}_{a}\eh^{N}_{b}\eta_{MN},
\end{align}
where $\eh^{M}_{a}$ are the boundary frame fields.\footnote{We choose $I,J=0,1,2,3$ for bulk and $M, N,\cdots=0,1,2$ to denote the tangent space directions on the boundary.} The frame fields allows us to write any boundary tensor as a collection of scalar fields, i.e. $X_{ab\cdots}=X_{MN\cdots}\eh^{M}_{a}\eh^{N}_{b}$. Taking this into account, the variation will only act on the boundary frame fields, while the tangent space components of $X_{MN\cdots}$ will be held fixed. In the case of asymptotically AdS spacetime, the choice of frame fields corresponds to the boundary metric in Fefferman-Graham expansion. 

This replacement of boundary metric with the boundary frame fields leads to the modified boundary stress tensor 
\begin{align}
T^{ab}\hat{\epsilon}=\frac{\delta S}{\delta \eh_{b}^{M}}\eh^{aM}.\label{emten3}
\end{align}
The two definitions \eqref{emten1} and \eqref{emten3} agree when the nontrivial boundary fields are metric and some scalars. However, if there are additional boundary fields, then \eqref{emten3} will be the correct one containing the contributions from the extras. In particular, consider the variation of a model supported by the non-abelian gauge fields with a relativistic conformal field theory at the boundary 
\begin{align}
\delta S=\int \hat{\epsilon} (T^{a}{}_{M}\delta\eh^{M}_{a}+s^{\Lambda}_{M}\delta A^{M}_{\Lambda}),\label{emten2}
\end{align}
where the covariant divergence of $T^{ab}$ can be shown to satisfy a modified conservation law
\begin{align}
\nabla_{a}T^{ab}=s^{\Lambda}_{M}\nabla^{b}A_{\Lambda}^{M}+T^{a}_{M}\nabla^{b}\eh^{M}_{a},
\end{align}
which can also be written as
\begin{align}
D_{a}T^{a}{}_{b}=s^{\Lambda}_{M}\partial_{b}A^{M}_{\Lambda} \,\,\text{with}\,\,D_{a}\eh^{M}_{b}=0.\label{conservation2}
\end{align}
This tensor allows the construction of the counter-term charges, which generate the correct asymptotic symmetries of the theory \cite{Hollands:2005ya}.

We now turn our attention to the non-relativistic limit of this prescription. In a non-relativistic field theory, instead of a covariant energy momentum tensor, we have an energy density $\mathcal{E}$, an energy flux $\mathcal{E}_i $, a momentum density $\mathcal{P}_i$ and a symmetric spatial stress tensor $\Pi_{ij}$. This \textit{stress tensor complex} satisfies the following conservation equations, i.e. diffeomorphism Ward identities
\begin{align}
\partial_{t}\mathcal{E}+\partial_{i}\mathcal{E}^{i}=0,\quad \partial_{t}\mathcal{P}_{j}+\partial_{i}\Pi^{i}{}_{j}=0.\label{diffid}
\end{align}
In addition to these, we have the dilatation Ward identity for theories with Lifshitz scaling symmetry
\begin{align}
z\mathcal{E}=\Pi^{i}{}_{i}.
\end{align}
The main idea behind the procedure of \cite{Ross:2009ar} is to consider the non-relativistic limit of the energy-momentum tensor (\ref{emten2}) since, all of the above identities can be derived as a non-relativistic limit of a relativistic conservation equation. For a consistent limit, the boundary frame fields should be related to the bulk frame fields by an appropriate power of $r$ such that $\eh^{M}_{b}$ stays finite as $r\rightarrow\infty$. With the following rescaling
\begin{align}
e^{(0)}=r^{z}\eh^{(0)},\quad e^{(i)}=r\eh^{(i)},\quad e^{(3)}=\frac{dr}{r},
\end{align}
the Lifshitz background solution is then the condition $\eh^{(0)}\rightarrow dr$, $\eh^{(i)}\rightarrow dx^{i}$, $A^{\Lambda}_{M}\rightarrow\sigma\delta_{M}^{\Lambda}$ as $r\rightarrow\infty$. Now, the scaled frame fields can be considered as the boundary data as follows
\begin{align}
\delta S= \int d^{3}y (-2s^{ab}\eh_{bM}+s^{a}_{\Lambda} A_{M}^{\Lambda})\delta\eh^{M}_{a}.\label{genvar2}
\end{align}
Decomposing \eqref{genvar2} into components of the stress energy complex gives
\begin{align}
\delta S=\int d^{3}y\left[-\mathcal{E}\delta\eh_{t}^{0}-\mathcal{E}^{i}\delta \eh_{i}^{0}
+\mathcal{P}_{i}\delta \eh_{t}^{i}+\Pi_{i}{}^{j}\delta\eh_{j}^{i}\right],
\end{align}
where 
\begin{align}
\mathcal{E}&=2s^{t}_{t}-s^{t}_{\Lambda}A_{t}^{\Lambda},&\mathcal{E}^{i}&=2s^{i}{}_{t}-s^{i}_{\Lambda}A_{t}^{\Lambda},\label{firstTset}\\
\mathcal{P}_{i}&=-2 s^{t}{}_{i}+s^{t}_{\Lambda}A_{i}^{\Lambda},&\Pi_{i}{}^{j}&=-2s^{j}{}_{i}+s^{j}_{\Lambda}A_{i}^{\Lambda}.\label{secTset}
\end{align}
To be precise, it is clear from the variation \eqref{genvar2} that the lower indices for $\mathcal{P}_{i}$ and $\Pi_{i}{}^{j}$ are flat. However, for the sake of simplicity, we converted all of them to the spacetime ones by multiplying the expressions with their respective frame fields.

To sum up, starting from the relativistic counterpart, we have reviewed the procedure of constructing a non-relativistic boundary stress tensor. Since the theory we consider involves non-scalar boundary fields, special attention must be given to their variation. This is handled by introducing vielbeins, which makes it possible to keep the components of boundary fields fixed, resulting in a stress tensor that has correct conserved charges. For the non-relativistic limit, the frame fields in Lifshitz spacetime need to be scaled appropriately to have finite values at the boundary. This approach is also intrinsically natural for non-relativistic theories where space and time are scaled differently.

There remains now to perform a perturbation analysis in order to check whether the stress tensor and the action are finite for a more general class of asymptotically Lifshitz spacetimes.

\section{Perturbation analysis}{\label{PERT}}
We now focus on a more general class of asymptotically Lifshitz spacetimes and show both the action and the stress tensor stays finite. As with most of the EYM solutions, the Lifshitz black holes are numerical solutions and studied in \cite{Devecioglu:2014iia}. We will return to their analysis in section \ref{numblackhole}. On the other hand, the exact solutions found in \cite{Fan:2015yza} are dressed with an extra Maxwell field and admit the colored Lifshitz background as a vacuum. We refrain ourselves studying these exact solutions with extra matter fields and resort to perturbative analysis with the perturbations that are asymptotic to the Lifshitz background. 

In the following, we first introduce the appropriate choice of basis for boundary frame fields, which allows the decomposition of the gauge field perturbation into scalar and vector components that respect the symmetry of the background solution. We then solve the linearised system for constant perturbations and identify the modes that preserve asymptotically Lifshitz condition. Employing these modes, we compute the on shell value of the action and components of the stress tensor, showing they are finite and obey Ward identities. In the final part of this section, we perform a similar analysis for the generalised perturbations with an extra step of expanding the linearised equations in powers of wave number and frequency in order to find solutions.

\subsection{Perturbation setup}{\label{pertsetup}}
Since we will perform the analysis in the linearised regime, let us carefully put together all the ingredients. We consider a generic metric as a background and its perturbation as follows
\begin{align}
g_{\mu\nu}=\bar{g}_{\mu\nu}+h_{\mu\nu},
\end{align}
where the deviation $h_{\mu\nu}$ should vanish sufficiently rapidly so that the metric approaches Lifshitz vacuum $\bar{g}_{\mu\nu}$ as $r\rightarrow\infty$. We will work in Gaussian gauge, i.e. $h_{\mu r}=0$ and the perturbations are also scaled by an appropriate power of $r$ accordant with the arguments in the previous section
\begin{align}
h_{tt}=-r^{2z}\hat{h}_{tt},\quad h_{ti}=-r^{2z} \vh_{1i}+r^{2} \vh_{2i},\quad h_{ij}= r^{2}\hat{h}_{ij}.
\end{align}
For this choice of perturbation, the orthonormal frames are\footnote{The background metric $\bar{g}_{\mu\nu}$ is responsible for raising and lowering of the spacetime indices $\mu,\nu\cdots$ and defining covariant derivative $\bar{\nabla}_{\mu}$. For the sake of simplicity, we also use the following shorthands: $\hh^{i}_{j}=\delta^{ik}\hh_{kj},\, \hh^{i}_{i}=\hh_{ij}\delta^{ij}$. }
\begin{align}
e^{(0)}=r^{z}\hat{e}^{(0)}=r^{z}\left((1+\frac{1}{2}\hat{h}_{tt})dt+\vh_{1i}dx^{i}\right),
\,\, e^{(i)}=r \hat{e}^{(i)}=r\left(\vh_{2i}dt+(\delta^{i}_{j}+\frac{1}{2}\hat{h}^{i}{}_{j})dx^{j}\right),\,\, e^{(3)}=\frac{dr}{r},
\end{align}
where asymptotically vanishing perturbations will lead to the Lifshitz background \eqref{backmet}. The metric determinants for the bulk and the boundary read respectively
\begin{align}
\sqrt{-g}=r^{z+1}\left[1+\frac{1}{2}\left(\hat{h}_{tt}+\hat{h}^{i}{}_{i}\right)\right],\quad \sqrt{-\gamma}=r^{z+2}\left[1+\frac{1}{2}\left(\hat{h}_{tt}+\hat{h}^{i}{}_{i}\right)\right].
\end{align}
On the matter sector, we define the first order gauge field perturbations as
\begin{align}
A^{\Lambda}_{\mu}=\bar{A}^{\Lambda}_{\mu}+a^{\Lambda}_{\mu},
\end{align}
where $\bar{A}_{\mu}$ is the background solution \eqref{backsol}
\begin{align} 
\bar{A}=\sigma r (T^{1}dx_{1}+T^{2}dx_{2}).
\end{align}

In order to decompose the field equations into vector and scalar parts, one must choose the gauge field perturbation $a_{\mu}$ such that it respects the symmetries of the background solution, i.e. a diagonal combination of the spatial rotation and an internal flavour symmetry rotating $A_{i}^{1}, A_{i}^{2}$ into each other\footnote{We are grateful to Simon F. Ross for the discussion of several issues in this section}. We will also keep all the components of the perturbation and let the field equations indicate the ones that are pure gauge, and therefore, can be set to zero. With these in mind, we make the following choice
\begin{align}
a_{\mu}dx^{\mu}=(r\sigma \vh_{2i}T^{i}-r^{z}\sigma \hat{a}_{3}T^{3})dt+\frac{1}{2}r\sigma\left(2\,\hat{b}_{i \Lambda}T^{\Lambda}
+\hat{h}_{ij}\delta^{j}_{k}T^{k}\right)dx^{i}+ \frac{\sigma}{r}\hat{c}_{\Lambda}T^{\Lambda}dr,\label{gaugeans}
\end{align}
where
\begin{align}
\bh_{i\Lambda}=\bh_{\Lambda i},\quad  \text{for}\quad \Lambda=1,2.
\end{align}
or equivalently $\bh_{ij}=\bh_{ji}$ which corresponds to mixing of spatial and group indices. The choice \eqref{gaugeans}  might seem quite arbitrary, but when written on flat indices 
\begin{align}
A^{I}_{\Lambda}T^{\Lambda}=\sigma\left(\hat{a}_{3}T^{3}\delta^{I}_{0}+\delta^{I}_{j}T^{j}+\hat{b}_{j\Lambda}T^{\Lambda}\delta^{Ij}+\hat{c}_{\Lambda}T^{\Lambda}\delta^{I}_{3}\right),\label{gaugeflat}
\end{align}
the compatibility with the prescription we gave is evident, i.e. for vanishing perturbations, the only non-zero component is the background solution which is constant on tangent directions. This choice is also quite tractable in terms of solving the coupled equations.

Finally, for completeness, we write down the linearized field equations for both the metric and the gauge field 
\begin{align}
\mathcal{E}_{\mu\nu}^{L}=&R_{\mu\nu}^{L}-\frac{1}{2}\bar{R}h_{\mu\nu}-\frac{1}{2}\bar{g}_{\mu\nu}R^{L}+\Lambda h_{\mu\nu}
-\dfrac{1}{g_{\text{\tiny YM}}^{2}}\text{Tr}\Big[2f_{(\mu}{}^{\alpha}\bar{F}_{\nu)\alpha}-\bar{F}_{\mu\sigma}\bar{F}_{\nu\alpha}h^{\sigma\alpha}\nonumber\\
&-\frac{1}{2}f_{\alpha\beta}\bar{F}^{\alpha\beta}\bar{g}_{\mu\nu}+\frac{1}{2}\bar{F}_{\alpha}{}^{\gamma}\bar{F}^{\alpha\beta}h_{\gamma\beta}\bar{g}_{\mu\nu}-\frac{1}{4}\bar{F}_{\alpha\beta}\bar{F}^{\alpha\beta}h_{\mu\nu}\Big],\label{einlin}\\
\mathcal{E}_{\nu}^{L}=&-\nb_{\alpha}f_{\nu}{}^{\alpha}+\nb_{\beta}\Fb_{\nu\alpha}h^{\alpha\beta}-\frac{1}{2}\Fb_{\nu}{}^{\alpha}\nb_{\alpha}h
+\Fb_{\nu}{}^{\alpha}\nb_{\beta}h_{\alpha}{}^{\beta}+\Fb_{\alpha\beta}\nb^{\beta}h_{\nu}{}^{\alpha}\nonumber\\
-& i \left( [\Ab_{\alpha}, f^{\alpha }_{\nu} ] +[ a_{\alpha} ,\Fb^{\alpha}_{\nu} ]
-[\Ab_{\alpha},\Fb^{\gamma}_{\nu}]h^{\alpha}{}_{\gamma} \right),\label{ymlin}
\end{align}
where $f_{\mu\nu}\equiv\bar{\nabla}_{\mu}a_{\nu}-\bar{\nabla}_{\nu}a_{\mu}$ and the remaining expressions for the linearised objects are given in the Appendix \ref{appendix1}.

Before discussing the solutions to the linearised equations, let us present the action and stress-energy complex in terms of perturbations. As a result of the background symmetry and the perturbation ansatz, the only Fourier modes that will contribute to the action after the integration over the boundary coordinates are the scalar ones
\begin{multline}
S=-\int^{r_{c}} dr\,r^{z+1}\Big\{ 2z(z+2)+z(z+2)(\hat{h}_{tt}+\hat{h}^{i}{}_{i})+2(z^{2}+z-2)\hat{b}^{i}{}_{i}
+2(z-1)r\partial_{r}\hat{b}^{i}{}_{i}\\
+(2z+3)r\partial_{r}[\hat{h}_{tt}+\hat{h}^{i}{}_{i}]+r^{2}\partial_{r}^{2}[\hat{h}_{tt}+\hat{h}^{i}{}_{i}]\Big\}\\
+r^{2+z}\left[2z+2(z-1)\hat{b}^{i}{}_{i}+z(\hat{h}_{tt}+\hat{h}^{i}{}_{i})+r\partial_{r}[\hh_{tt}+\hat{h}^{i}{}_{i}]\right]\Big|_{r=r_{c}}.\label{actionlinear}
\end{multline}
This result also implies that there will be no contribution from a possible $S_{deriv}$ at linear order.  In the following subsection, we will find the explicit solutions for the perturbations and show \eqref{actionlinear} is finite.

On the other hand, in contrast to the action, all modes contribute to the stress tensor complex 
\begin{align}
\mathcal{E}&=-r^{z+2}(z-1)\left(2\hat{b}^{i}{}_{i}+\frac{r\partial_{r}\hat{h}^{i}{}_{i}}{(z-1)}-\frac{2\epsilon^{ij}\partial_{j}{\bh_{i3}}}{r\sqrt{z+1}}\right)+\tilde{\mathcal{E}},\label{energy}\\
\mathcal{E}^{i}&=-r^{z+2}(z-1)\delta^{ij}\left(\frac{r^{2z+1}\partial_{r}\hat{v}_{1j}}{(z-1)}-\frac{r\partial_{r}\vh_{2j}}{(z-1)}-\frac{2}{\sqrt{z+1}}\epsilon_{j}{}^{k}(r^{z}\partial_{k}\ah_{3}+r\partial_{t}\bh_{k3})\right)+\tilde{\mathcal{E}}^{i},\label{epsud}\\
\mathcal{P}_{i}&=-r^{z+2}(z-1)\left(\frac{r\partial_{r}\hat{v}_{1i}}{(z-1)}
-\frac{(2z-1)}{(z-1)}r^{-2z+3}\partial_{r}\vh_{2i}+2r^{1-2z}\partial_{t}\ch_{i}\right)+\tilde{\mathcal{P}}_{i},\label{momenta}\\
\Pi^{i}{}_{j}&=r^{z+2}(z-1)\Big(2(-2\bh^{i}{}_{j}+\delta^{i}{}_{j}\bh^{k}{}_{k})+2\sqrt{z+1}\epsilon^{i}{}_{j}\ch_{3}
+\frac{2}{r}\partial^{i}\ch_{j}-2r\partial_{r}\bh^{i}{}_{j}\nonumber\\
&+\frac{r}{(z-1)}\partial_{r}[\delta^{i}{}_{j}(\hh_{tt}+\hh^{k}{}_{k})-z\hh^{i}{}_{j}]\Big)+\tilde{\Pi}^{i}{}_{j},\label{spatial}
\end{align}
where quantities with tilde denote the possible $S_{deriv}$ contributions.

Several observations are in order here. Firstly, all components of the matter sector of the perturbation are multiplied by $(z-1)$, so in the limit $z=1$, they are all decoupled, and only the metric perturbations survive. The remaining piece should coincide with the relativistic definition \eqref{emten1}. Secondly, now it is possible to consider the $S_{deriv}$ contributions from the action, which were quite important to make the components of $\Pi^{i}{}_{j}$ finite for generalised perturbations in \cite{Ross:2009ar}. Thirdly, although we discussed the importance of the modification of conservation equation \eqref{conservation2}, when linearised, the right-hand side of the equation reads
 \begin{align}
D_{a}T^{a}{}_{b}=\bar{s}_{M}^{\Lambda}(\partial_{b}A^{M}_{\Lambda})^{L}+(s_{M}^{\Lambda})^{L}\partial_{b}\bar{A}^{M}_{\Lambda}.
 \end{align}
The first term on the right-hand side vanishes on-shell and the second one vanishes from the constancy of the background gauge field on flat indices. Therefore, the linearisation is blind to this modification. Also, unlike its relativistic counterpart, the spatial energy-momentum tensor \eqref{spatial} by construction is not symmetric off-shell. However, we expect to have a symmetric one on-shell, or it can always be improved to be symmetric. Finally, it is worth repeating the argument of \cite{Ross:2009ar} on the applicability of the linearisation. The equations \eqref{energy}-\eqref{spatial} imply that the modes with a fall off $r^{-(z+2)}$ will contribute to the components of stress-energy tensor at linear order. Then, any possible solution with $r^{-(z+2)/2}$ fall off will also contribute to the quadratic order, signalling the insufficiency of the linear regime and a need for non-linear analysis.

We are now in a position to find the solutions to the perturbations and explicitly compute the components of the stress-energy complex, and show it satisfies the Ward identities.

\subsection{Constant perturbations and solutions}{\label{constantperturbations}}
Let us first consider the perturbations that are constant in the boundary directions. With the help of the symmetry of mixing the spatial and the flavour indices, we can decompose the constant perturbations into scalar $f(r),j(r),k(r)$, vector $\hat{v}_{1i}(r),\vh_{2i}(r)$ and tensor modes $t_{d}(r),t_{o}(r),b_{d}(r),b_{o}(r)$ as follows 
\begin{align}
h_{tt}=-r^{2z}f(r),\quad h_{ti}=-r^{2z}\vh_{1i}(r)+r^{2} \vh_{2i}(r),\quad h_{ij}=r^{2}k(r)\delta_{ij}+r^{2}k_{ij},\quad \bh_{ij}=\delta_{ij}j(r)+J_{ij},
\end{align}
where
\begin{align}
k_{ij}=\begin{pmatrix}
t_{d}(r)& t_{o}(r) \\
t_{o}(r) & -t_{d}(r)
\end{pmatrix},\quad
J_{ij}=\begin{pmatrix}
b_{d}(r) & b_{o}(r) \\
b_{o}(r) & -b_{d}(r)
\end{pmatrix}.
\end{align}
In addition to the vector modes analogous to \cite{Ross:2009ar}, we also have tensor modes for the gauge field perturbation.
\subsubsection{Scalar modes}
The field equations for the constant scalar perturbations read
\begin{align}
2 r^2 j''(r)&=-8 r j'(r)-r f'(r)+2 z (z+3) j(r),\\
r^2 f''(r)&=-2  (z+1)r f'(r)+2  (z-1) r j'(r)+2 \left(3 z^2+z-4\right) j(r),\\
2 r k'(r)&=-r f'(r)+2 (z-1) r j'(r)-2z (z-1)   j(r).
\end{align}
For $z\geq 1$, the solutions are
\begin{align}
j(r)=& \fc_{1}r^{-(z+2)}+\fc_{2}r^{-\frac{1}{2}(z+2+\beta_{z})}+\mathfrak{c}_{3}r^{-\frac{1}{2}(z+2-\beta_{z})},\label{jey}\\
f(r)=& -\frac{4\fc_{1}(z+1)}{z+2}r^{-(z+2)}+
\frac{2\fc_{2}(z-1)(-2+3z+\beta_{z})}{2+z+\beta_{z}}r^{-\frac{1}{2}(z+2+\beta_{z})}\nonumber\\
&+\frac{2\fc_{3}(z-1)(-2+3z-\beta_{z})}{2+z-\beta_{z}}r^{-\frac{1}{2}(z+2-\beta_{z})}+\fc_{4},\label{fe}\\
k(r)=&\frac{2\fc_{1}z(1+z)}{2 + z}r^{-(z+2)}
+\frac{4\fc_{2}(z-1)}{2+z+\beta_{z}}r^{-\frac{1}{2}(z+2+\beta_{z})}+\frac{4\fc_{3}(z-1)}{2+z-\beta_{z}}r^{-\frac{1}{2}(z+2-\beta_{z})}+\fc_{5},\label{key}
\end{align}
where $\beta_{z}\equiv\sqrt{9 z^2+4 z+12}$. In order to satisfy the asymptotically Lifshitz boundary conditions we should set constants $\fc_{4}=\fc_{5}=0$, which is merely a redefinition of coordinates. Moreover, $\fc_3$ should set to be zero since it is a growing mode as $r\rightarrow\infty$. Therefore, we are left with two free parameters $\fc_{1},\fc_{2}$ for the constant scalar sector. With the scalar mode solutions in hand, we can now calculate the on-shell value of the action \eqref{actionlinear} 
\begin{align}
S=-\frac{4\fc_{1}(z-1)z}{z+2}.
\end{align}
Energy \eqref{energy} is also composed of only scalar modes
\begin{align}
\mathcal{E}=-2r^{z+2}\left(2(z-1)j+rk^{\prime}\right)=4 \fc_{1}(z^{2}+1).\label{energyfin}
\end{align}
As a side note, even we let all the modes in the solution \eqref{jey}-\eqref{key} the on-shell action and the energy stays the same. 

Before moving on to the vector and tensor modes, let us present the results on the $\hat{c}_{\Lambda}(r)$ and $\bh_{i3}(r)$ components of the gauge field perturbation. The field equations for these components are
\begin{align}
(z+2)\hat{c}_{i}(r)+r\hat{c}^{\prime}_{i}(r)+\frac{r \epsilon_{i}{}^{j}}{\sqrt{z+1}}\left[(z+3)\hat{b}^{\prime}_{j3}(r)+r\hat{b}^{\prime\prime}_{j3}(r)\right]=&0,\\
\hat{c}_{i}(r)+\frac{r \epsilon_{i}{}^{j}\,\bh^{\prime}_{j3}(r)}{\sqrt{z+1}}=&0\label{cieqn}.
\end{align}
This system of equations is underdetermined, signalling the perturbations $\hat{c}_{i}$ and $\hat{b}_{i3}$ are pure gauge. In stress-energy tensor \eqref{energy}-\eqref{spatial}, these functions appear with time and spatial derivatives, therefore, do not contribute to the constant perturbations, maintaining the gauge independence of conserved quantities. Finally, $\ch_{3}(r)$ is fixed to zero by its field equation, ensuring a symmetric spatial stress tensor. 
\subsubsection{Vector modes}
We next consider the contribution from the vector modes. The perturbations $\vh_{1i}(r)$, $\vh_{2i}(r)$ are coupled 
\begin{align}
r^{2}\vh_{1i}''-r^{4-2z}\vh_{2i}''+(3z+1)r\vh_{1i}'-(z+3)r^{3-2z}\vh_{2i}'&=0,\\
r^{2}\vh_{2i}''+r^{2z-1}\vh_{1i}'-(z-4)r\vh_{2i}'&=0,
\end{align}
and $\ah_{3}(r)$ has its own decoupled equation
\begin{align}
r^{2}\ah_{3}^{\prime\prime}(r)&+(z+3)r\ah_{3}^{\prime}(r)-2\ah_{3}(r)=0.
\end{align}
The solutions of $\vh_{1i},\vh_{2i}$ has two branches, for $z\neq4$ 
\begin{align}
\vh_{1i}(r)&=-\frac{\fc_{1i}}{z+2}\,r^{-(z+2)}  + \frac{(1-2z)\fc_{2i}}{3z}\,r^{-3z}+\fc_{3i},\label{v1sol}\\
\hat{v}_{2i}(r)&=-\frac{\fc_{1i}}{z-4} r^{-(4-z)}-
  \frac{\fc_{2i}}{z+2} r^{-(z+2)} +  \fc_{4i},\label{v2sol}\\
\end{align}
and $z=4$
\begin{align}
\vh_{1i}(r)&=-\frac{\fc_{1i}}{6r^{6}}-\frac{7\fc_{2i}}{12r^{12}}+\fc_{3i},\\
\vh_{2i}(r)&=\fc_{1i}\log{r}-\frac{\fc_{2i}}{6r^{6}}+\fc_{4i}.
\end{align}
Finally the solutions for $\ah_{3}(r)$ reads
\begin{align}
\ah_{3}(r)=&\mathfrak{a}_{31}\,r^{\frac{-(2 \sqrt{2} +\sqrt{2} z + \sqrt{24 + 8 z + 2 z^2})}{2\sqrt{2}}}
+\mathfrak{a}_{32}\,r^{\frac{-(2 \sqrt{2} +\sqrt{2} z - \sqrt{24 + 8 z + 2 z^2})}{2\sqrt{2}}}.
\end{align}
From the solutions we compute the energy flux and the momentum density as follows
\begin{align}
\mathcal{E}^{i}=&-2\fc_{2j}\delta^{ij}(z-1),\label{eflux}\\
\mathcal{P}_{i}=&2\fc_{1i}(z-1).\label{momden}
\end{align}
The solutions are similar to the ones in \cite{Ross:2009ar} so their analysis is the same. The coefficient $\fc_{3i}$ represents the shifting of the coordinate $t\rightarrow t+\fc_{3i} x^{i}$, and the coefficient $\fc_{4i}$ represents the shift $x^{i}\rightarrow x^{i}+\fc_{4i}t$, therefore they are a pure gauge and can be set to zero. The linearisation is adequate for $\fc_{1i}$ in $\vh_{2i}$ only for $2>z>1$ and at $z\geq 4$, we need to set it to zero since it becomes a growing mode, spoiling the boundary conditions. 

Finally, the solution $\ah_{3}(r)$ corresponds to the perturbations in the electric charge. Since the background solution is purely magnetic, this mode does not contribute to the stress-energy complex as it appears with a boundary coordinate derivative in \eqref{epsud}. Therefore, we will ignore this solution in constant perturbations. 

\subsubsection{Tensor modes}
There are two coupled systems of equations for tensor modes. The first one involves the diagonal components
\begin{align}
2 r^2 b_{d}''(r)+r^2 t_{d}''(r)+r (z+1)t_{d}'(r)+2 r (z+3) b_{d}'(r)+4 (z+1) b_{d}(r)=&0,\\
r^2 t_{d}''(r)+r (3 z+1) t_{d}'(r)+4 r (z-1) b_{d}'(r)+4 (z-1)b_{d}(r)=&0,
\end{align}
with the solutions
\begin{align}
t_{d}(r)=&
   \ft_{d1}\,r^{-(z+2)} + \ft_{d2}\,r^{-\frac{1}{2}(z+2+\xi_{z})}+
   \ft_{d3}\,r^{-\frac{1}{2}(z+2-\xi_{z})}+\ft_{d4},\label{td}\\
   b_{d}(r)=&-\frac{\ft_{d1}(z+2)  }{2 (z+1)}r^{-(z+2)}
   + \frac{\ft_{d2} (2-3 z+\xi_{z})}{4 (z-1)}r^{-\frac{1}{2}(z+2+\xi_{z})}
   + \frac{\ft_{d3} (2-3 z-\xi_{z})}{4 (z-1)}r^{-\frac{1}{2}(z+2-\xi_{z})},\label{bd}
\end{align}
where $\xi_{z}\equiv \sqrt{z^2-12 z+4}$. The parameter $\xi_{z}$ is real for $z\geq 2(3+2\sqrt{2})$. In this region it is possible to include the mode $\ft_{d2}$. However, its sister mode $\ft_{d1}$ falls slower than $r^{-(z+2)/2}$ so again linear analysis is insufficient to understand this mode clearly. The constant term $\ft_{d4}$ corresponds to the relative scaling of the $x,y$ coordinates and can be ignored.

The second system is for the off-diagonal tensor modes which have the same structure
\begin{align}
2 r^2 b_{o}''(r)+r^2 t_{o}''(r)+r (z+1)t_{o}'(r)+2 r (z+3) b_{o}'(r)+4 (z+1) b_{o}(r)=&0,\\
r^2 t_{o}''(r)+r (3 z+1) t_{o}'(r)+4 r (z-1) b_{o}'(r)+4 (z-1)b_{o}(r)=&0,
\end{align}
where the solutions read
\begin{align}
t_{o}(r)=&
   \ft_{o1}\,r^{-(z+2)} + \ft_{o2}\,r^{-\frac{1}{2}(z+2+\xi_{z})}+
   \ft_{o3}\,r^{-\frac{1}{2}(z+2-\xi_{z})}+\ft_{o4},\label{to}\\
 b_{o}(r)=&-\frac{\ft_{o1} (z+2) }{2 (z+1)}r^{-(z+2)}+\frac{\ft_{o2} (2-3 z+\xi_{z} ) }{4 (z-1)}r^{-\frac{1}{2}(z+2+\xi_{z})}+\frac{\ft_{o3} ( 2-3 z-\xi_{z} ) }{4 (z-1)}r^{-\frac{1}{2}(z+2-\xi_{z})}.\label{bo}
\end{align}
The $z$ bound on $\ft_{o2}$ and the fall rate argument of $\ft_{o3}$ is identical. This time, the pure gauge mode $\ft_{o4}$ is related to the rotation of $x,y$ coordinates. 

Now, armed with the tensor solutions, we can finally work out the spatial stress tensor and check the scaling Ward identity. The components of $\Pi^{i}{}_{j}$ in terms of decomposition functions are as follows
\begin{align}
\Pi^{1}{}_{1}=&-r^{z+2} \Big(4 (z-1)b_{d}+r \frac{d}{dr}[2 (z-1) b_{d}-f+2 (z-1) j+(z-2) k+z t_{d}]\Big),\\
\Pi^{2}{}_{2}=&r^{z+2} \Big(4 (z-1)b_{d}+r \frac{d}{dr}[2 (z-1) b_{d}+f-2 (z-1) j-(z-2) k+z t_{d}]\Big),\\
\Pi^{1}{}_{2}=&-r^{z+2} \Big(2 r (z-1) b_{o}^{\prime}+4 (z-1) b_{o}+r z t_{o}^{\prime}\Big).
\end{align}
Plugging in the solutions we have
\begin{align}
\Pi^{1}{}_{1}=&2 \fc_{1} z\left(z^2+1\right)+\frac{2 \ft_{d1} z (z+2)}{z+1}, \\
\Pi^{2}{}_{2}=&2 \fc_{1} z\left(z^2+1\right)-\frac{2 \ft_{d1} z (z+2)}{z+1},\\
\Pi^{1}{}_{2}=&\Pi^{2}{}_{1}=\frac{2 \ft_{o1} z (z+2)}{z+1}.
\end{align}
Taking these components and the energy \eqref{energyfin} into account, we see the relation $z\mathcal{E}=\Pi^{i}{}_{i}$ is satisfied. However, a non-trivial check for the conservation equations \eqref{diffid} is not possible here since all the components of the stress-energy complex are constant, leading to a trivial result. Therefore, we postpone that check to the next section, where we study generalised perturbations. Finally, as we have checked all the solutions, it is now easy to see that a smarter choice for the perturbative part of the gauge field is 
\begin{align}
a_{\mu}dx^{\mu}=r\sigma \vh_{2i}T^{i}dt+\frac{1}{2}r\sigma\left(2\,\hat{b}_{ij}T^{j}
+\hat{h}_{ij}\delta^{j}_{k}T^{k}\right)dx^{i},\label{gaugeans2}
\end{align}
or in flat indices
\begin{align}
A^{I}_{\Lambda}T^{\Lambda}=\sigma\left(\delta^{I}_{i}+\hat{b}_{ij}\delta^{Ij}\right)T^{i}.
\end{align}
\subsection{General perturbations and solutions}{\label{generalpert}}
Having dealt with the constant perturbations, let us switch the gears and continue with the perturbations that depend on time and boundary coordinates in the form of plane waves along $x$ direction. Again, exploiting the background symmetry, we will group the perturbations as scalar and vector parts. We also introduce factors of $k$ and $\omega$ to ensure that the field equations will involve even powers of $k$, $\omega$ so that we can perform an expansion, as the system of equations will get quite complicated to solve in full form.

With this in mind, the scalar perturbations for the metric side read
\begin{align}
\hh_{tt}=&f(r)e^{i(\omega t+kx)}, &\vh_{11}&=k\,\omega s_{1}(r) e^{i(\omega t+kx)},\\
\quad \hh_{xx}=&(k_{L}(r)+k^{2}k_{T}(r))e^{i(\omega t+kx)},& \hh_{yy}&=(k_{L}(r)-k^{2}k_{T}(r))e^{i(\omega t+kx)},
\end{align}
and on the matter side we have
\begin{align}
\bh_{11}=&(b_{L}(r)+k^{2}b_{T}(r))e^{i(\omega t+kx)},& \bh_{22}&=(b_{L}(r)-k^{2}b_{T}(r))e^{i(\omega t+kx)},\quad\\
\vh_{21}=& k\,\omega s_{2}(r)e^{i(\omega t+kx)},&\bh_{23}&=ik s_{3}(r)e^{i(\omega t+kx)},\\
\ch_{1}=& ik s_{4}(r)e^{i(\omega t+kx)}.
\end{align}
On the other hand, the vector perturbations for the metric are as follows
\begin{align}
\vh_{12}&={} k \, \omega \mathsf{v}_{1}(r)e^{i(\omega t+k x)},\\
\hh_{12}&={}\mathsf{v}_{2}(r)e^{i(\omega t+k x)},
\end{align}
and finally the matter perturbations are chosen as
\begin{align}
\vh_{22}&=k\,\omega \mathsf{v}_{3}(r)e^{i(\omega t+k x)}, & \bh_{13}&= i k \mathsf{v}_{6}(r)e^{i(\omega t+k x)}, \\
\ah_{3}&= i\omega \mathsf{v}_{4}(r)e^{i(\omega t+k x)},& \ch_{2}&=i k \mathsf{v}_{7}(r)  e^{i(\omega t+k x)},\\
\bh_{12}&= \mathsf{v}_{5}(r)e^{i(\omega t+k x)}, &\ch_{3}&= \mathsf{v}_{8}(r)  e^{i(\omega t+k x)}.
\end{align}
In the coming sections, we will find solutions to these perturbations starting from the scalar modes and prove that the conservation laws are satisfied.
\subsubsection{Scalar modes}
\begingroup
\allowdisplaybreaks
Let us start with the scalar mode functions that contribute to the stress energy tensor complex in the following way
\begin{align}
\mathcal{E}=&-2 r^{z+2} \left[2(z-1)b_{L}+r k_{L}^{\prime}-\frac{k^{2}(z-1)s_{3}}{r\sqrt{z+1}} \right]e^{i(\omega t+k x)},\label{genen}\\
\mathcal{E}^{x}=& -k\omega r^{z+2}\left[r^{2z-1}s_{1}^{\prime}-rs_{2}'+\frac{2(z-1)s_{3}}{r\sqrt{z+1}}\right] e^{i(\omega t+k x)},\label{enden}\\
\mathcal{P}_{x}=&-k\omega r^{z+2}\left[rs_{1}^{\prime}-(z-1)r^{-2z+3}s_{2}'-2(z-1)r^{1-2z}s_{4}\right]e^{i(\omega t+k x)}\\
\Pi^{x}{}_{x}=&-r^{z+2}\Bigg[k^{2}\left\{4(z-1)b_{T}+2(z-1)rb_{T}^{\prime}+z r k_{T}^{\prime}+\frac{2(z-1)s_{4}}{r}\right\}\nonumber\\
&+r(2(z-1)b_{L}^{\prime}-f^{\prime}+(z-2)k_{L}^{\prime})\Bigg]e^{i(\omega t+k x)},\\
\Pi^{y}{}_{y}=&-r^{z+2}\Bigg[-k^{2}\left\{4(z-1)b_{T}+2(z-1)rb_{T}^{\prime}+z r k_{T}^{\prime}\right\}\nonumber\\
&+r(2(z-1)b_{L}^{\prime}-f^{\prime}+(z-2)k_{L}^{\prime})\Bigg]e^{i(\omega t+k x)}.
\end{align}
Plugging in the plane wave ansatz, the conservation equations \eqref{diffid} will yield
\begin{align}
\omega \mathcal{E}+k\mathcal{E}^{x}=0,\quad \omega \mathcal{P}_{x}+k\Pi^{x}{}_{x}=0,\quad z\mathcal{E}=&\Pi^{i}{}_{i}. \label{expcons}
\end{align}
It can be shown that the first conservation equation is equal to $tr$ component of the linearized Einstein equations \eqref{einlin}, i.e. $\mathcal{E}_{tr}$, and the second one is a combination of the $\mathcal{E}_{x_{1}r}$ and $\mathcal{E}_{r}^{1}$, therefore the conservation laws are guaranteed to be satisfied on-shell.

In this system of equations, there are nine functions to solve, and there are twelve equations in total, so three of the equations can be written as a linear combination of others. In Appendix \ref{appendix2}, we give all the field equations and the relations in detail. The main difference from the constant perturbations is that we lose the Euler type equation structure. Thus, we are not able to solve the system in its full form, and our line of attack will be expanding the perturbation functions in even powers of $k,\omega$, i.e. $F=\sum_{m,n}k^{2m}\omega^{2n}F^{(m,n)}$ where $m,n=0,1,2\cdots$. Then, the components of the stress energy tensor can be written as
\begin{align}
\mathcal{E}=&\sum_{m,n}k^{2m}\omega^{2n}(\mathcal{E}_{1}^{(m,n)}+k^{2}\mathcal{E}_{2}^{(m,n)})e^{i(\omega t+k x)},\label{Eexp}\\
\mathcal{E}^{x}=&\sum_{m,n}k^{2m+1}\omega^{2n+1}\mathcal{E}^{x(m,n)}e^{i(\omega t+k x)},\label{Exexp}\\
\mathcal{P}_{x}=&\sum_{m,n}k^{2m+1}\omega^{2n+1}\mathcal{P}^{(m,n)}_{x}e^{i(\omega t+k x)}\label{Pxexp}\\
\Pi^{x}{}_{x}=&\sum_{m,n}k^{2m}\omega^{2n}(\Pi^{(m,n)}_{L}+k^{2}\Pi^{(m,n)}_{S}+k^{2}\Pi^{(m,n)}_{T})e^{i(\omega t+k x)},\label{Pxxexp}\\
\Pi^{y}{}_{y}=&\sum_{m,n}k^{2m}\omega^{2n}(\Pi^{(m,n)}_{L}-k^{2}\Pi^{(m,n)}_{T})e^{i(\omega t+k x)},\label{Pyyexp}
\end{align}
where the following are defined to evaluate \eqref{expcons} in a more organized way
\begin{align}
\mathcal{E}_{1}^{(m,n)}=&-2r^{z+2}\left[2(z-1)b_{L}^{(m,n)}+rk_{L}^{(m,n)\prime}\right],\label{E1comp}\\
\mathcal{E}_{2}^{(m,n)}=& -2r^{z+2}\left[-\frac{(z-1)s_{3}^{(m,n)}}{r\sqrt{z+1}} \right],\\
\mathcal{E}^{x(m,n)}=&-r^{z+2}\left[r^{2z-1}s_{1}^{(m,n)\prime}-rs_{2}^{(m,n)\prime}+\frac{2(z-1)s_{3}^{(m,n)}}{r\sqrt{z+1}}\right],\label{Excomp}\\
\mathcal{P}_{x}^{(m,n)}=&-r^{z+2}\left[rs_{1}^{(m,n)\prime}-(z-1)r^{-2z+3}s_{2}^{(m,n)\prime}-2(z-1)r^{1-2z}s_{4}^{(m,n)}\right],\label{Pxcomp}\\
\Pi^{(m,n)}_{L}=&-r^{z+2}\left[2(z-1)rb_{L}^{(m,n)\prime}-rf^{(m,n)\prime}+(z-2)rk_{L}^{(m,n)\prime}\right],\label{PiLcomp}\\
\Pi^{(m,n)}_{T}=&-r^{z+2}\left[4(z-1)b_{T}^{(m,n)}+2(z-1)rb_{T}^{(m,n)\prime}+zrk_{T}^{(m,n)\prime}\right],\label{PiTcomp}\\
\Pi^{(m,n)}_{S}=&-r^{z+2}\left[\frac{2(z-1)s_{4}^{(m,n)}}{r}\right]. \label{PiScomp}
\end{align}
Before moving on to the solutions, let us impose conservation equations \eqref{expcons} to \eqref{Exexp}-\eqref{Pyyexp} and infer at which order we would get non-zero contribution for the components \eqref{E1comp}-\eqref{PiTcomp}. 
\begin{align}
\mathcal{E}_{1}^{(m+1,n)}+\mathcal{E}_{2}^{(m,n)}+\mathcal{E}^{x(m,n)}&=0,\label{finconsv1}\\
\mathcal{P}_{x}^{(m+1,n)}+\Pi^{(m+1,n+1)}_{L}+\Pi^{(m,n+1)}_{T}+\Pi^{(m,n+1)}_{S}&=0,\label{finconsv2}\\
z(\mathcal{E}_{1}^{(m+1,n)}+\mathcal{E}_{2}^{(m,n)})-2\Pi^{(m+1,n)}_{L}-\Pi^{(m,n)}_{L}&=0.\label{finconsv3}
\end{align}
From the first equation it follows that $\mathcal{E}^{(0,n)}_{1}=0$. Then, using second and third relations, it is easy to see $\Pi^{(0,n)}_{L}=\mathcal{P}_{x}^{(0,n)}=0$.

Close scrutiny of the expanded field equations will reveal the following structure: at every order $(m,n)$ there will be a part of field equations that will correspond to the homogeneous solutions which are the same with constant perturbations. Then through the expansion, these homogeneous solutions are sourced by lower-order expansion ones, leading to the particular integrals. However, the particular integrals are always suppressed by powers of $r$, therefore not contributing to the stress energy tensor. As a simple represantative of this structure, consider the following $(0,0)$ and $(1,0)$ order expansions of the equation \eqref{eq:1}
\begin{align}
r^{2}\left(k_{L}^{(0,0)\prime\prime}+f^{(0,0)\prime\prime}\right)+r\left[(z+3)k_{L}^{(0,0)\prime}+2(z+1)f^{(0,0)\prime}\right]-4(z^{2}-1)b_{L}^{(0,0)}=&0\\
r^{2}\left(k_{L}^{(1,0)\prime\prime}+f^{(1,0)\prime\prime}\right)+r\left[(z+3)k_{L}^{(1,0)\prime}+2(z+1)f^{(1,0)\prime}\right]-4(z^{2}-1)b_{L}^{(1,0)}=&\frac{f^{(0,0)}}{2r^{2}}-\frac{2(z^{2}-1)}{\sqrt{z+1}}\frac{s_{3}^{(0,0)}}{r}.
\end{align}
Therefore, once the zeroth-order equations are solved, the first order homogenous solutions are identical with extra particular solutions sourced by the zeroth-order ones. Moreover, unlike constant perturbations, the solutions now have extra constraints which will relate different modes to ensure the conservation equations are satisfied. 

Taking all these into account, we have the following solution structure at the order $(m,n)$ for the equations \footnote{In this and the following section, we are going to use the boldfaced characters for labelling modes, e.g. $\mathbf{k}^{\text{\tiny 103}}_{\text{\tiny T}}$ where first two upper indices tags the corresponding expansion parameter for wavenumber $k$ and frequency $\omega$ respectively, i.e. $k=1$ and $\omega=0$ in this case. The last index is the mode number.}

For $b_{L}^{(m,n)},k_{L}^{(m,n)},f^{(m,n)}$
\begin{align}
b_{L}^{(m,n)}(r)&=\mathbf{b}^{\text{\tiny mn1}}_{\text{\tiny L}}\, r^{-(z+2)}
+\mathbf{b}^{\text{\tiny mn2}}_{\text{\tiny L}}\, r^{-\frac{1}{2}(z+2+\beta_{z})}+\mathcal{O}(r^{p})\label{solbL}\\
f^{(m,n)}(r)&=\mathbf{f}^{\text{\tiny mn1}}\, r^{-(z+2)}
+\mathbf{f}^{\text{\tiny mn2}}\, r^{-\frac{1}{2}(z+2+\beta_{z})}+\mathcal{O}(r^{p})\label{solf}\\
k_{L}^{(m,n)}(r)&=\mathbf{k}^{\text{\tiny mn1}}_{\text{\tiny L}}\, r^{-(z+2)}
+\mathbf{k}^{\text{\tiny mn2}}_{\text{\tiny L}}\, r^{-\frac{1}{2}(z+2+\beta_{z})}+\mathcal{O}(r^{p})\label{solkL}
\end{align}
for $k_{T}^{(m,n)},b_{T}^{(m,n)}$
\begin{align}
k_{T}^{(m,n)}(r)&=\mathbf{k}^{\text{\tiny mn1}}_{\text{\tiny T}}\,r^{-(z+2)}+\mathbf{k}^{\text{\tiny mn2}}_{\text{\tiny T}}\,r^{-\frac{1}{2}(z+2-\xi_{z})}+\mathbf{k}^{\text{\tiny mn3}}_{\text{\tiny T}}\,r^{-\frac{1}{2}(z+2+\xi_{z})}+\mathcal{O}(r^{p}),\label{kT001eqn}\\
b_{T}^{(m,n)}(r)&=\mathbf{b}^{\text{\tiny mn1}}_{\text{\tiny T}} \,r^{-(z+2)}
+\mathbf{b}^{\text{\tiny mn2}}_{\text{\tiny T}} \, r^{-\frac{1}{2}(z+2-\xi_{z})}
+\mathbf{b}^{\text{\tiny mn3}}_{\text{\tiny T}} \,r^{-\frac{1}{2}(z+2+\xi_{z})}+\mathcal{O}(r^{p}).\label{bT001eqn}
\end{align}
for $s_{2}^{(m,n)}(r),s_{1}^{(m,n)}(r)$
\begin{align}
s_{1}^{(m,n)}(r)&=\mathbf{s}^{\text{\tiny mn1}}_{\text{\tiny 1}}\,r^{-(z+2)}
+\mathbf{s}^{\text{\tiny mn2}}_{\text{\tiny 1}}\,r^{-3z}+\mathcal{O}(r^{p})\\
s_{2}^{(m,n)}(r)&=\mathbf{s}^{\text{\tiny mn1}}_{\text{\tiny 2}}\, r^{-(z+2)}
+\mathbf{s}^{\text{\tiny mn2}}_{\text{\tiny 2}}\,r^{-(4-z)}+\mathcal{O}(r^{p})
\end{align}
for $s_{3}^{(m,n)}(r),s_{4}^{(m,n)}(r)$
\begin{align}
s_{3}^{(m,n)}(r)&=\mathbf{s}^{\text{\tiny mn1}}_{\text{\tiny 3}}\, r^{-(z+1)}+\mathcal{O}(r^{p})\\
s_{4}^{(m,n)}(r)&=\mathbf{s}^{\text{\tiny mn1}}_{\text{\tiny 4}}\,r^{-(z+1)}+\mathcal{O}(r^{p})\label{sols4}
\end{align}
where $\mathcal{O}(r^{p})$ corresponds to particular integrals that do not contribute to the stress energy tensor.  Actually we will only track down and solve the modes which contribute to the stress energy complex when evaluated at infinity. To that end, ''solving'' here will mean finding out the relation between the contributing modes. After meticulousy working out the expanded field equations, we have the following solutions
\begin{align*}
\mathbf{f}^{\text{\tiny 0n1}}&=0,\quad n\geq 0,\\
\mathbf{f}^{\text{\tiny mn1}}&=-\frac{4(z+1)\mathbf{b}^{\text{\tiny mn1}}_{\text{\tiny L}}}{z+2},\quad  m> 0, n\geq 0,\\
\mathbf{k}^{\text{\tiny 0n1}}_{\text{\tiny L}}&=0,\quad n\geq 0,\\
\mathbf{k}^{\text{\tiny mn1}}_{\text{\tiny L}}&=\frac{2z(z+1)\mathbf{b}^{\text{\tiny mn1}}_{\text{\tiny L}}}{z+2},\quad m>0,n\geq 0,\\
\mathbf{b}^{\text{\tiny 0n1}}_{\text{\tiny L}}&=0,\quad n\geq 0,\\
\mathbf{b}^{\text{\tiny m01}}_{\text{\tiny T}}&=\frac{1}{2}(z^{2}+1)\mathbf{b}^{\text{\tiny (m+1)01}}_{\text{\tiny L}},\quad m\geq 0,\\
\mathbf{k}^{\text{\tiny m01}}_{\text{\tiny T}}&=-\frac{(z+1)(z^{2}+1)\mathbf{b}^{\text{\tiny (m+1)01}}_{\text{\tiny L}}}{z+2},\quad m\geq 0,\\
\mathbf{k}^{\text{\tiny mn1}}_{\text{\tiny T}}&=-\frac{2(z+1)\mathbf{b}^{\text{\tiny mn1}}_{\text{\tiny T}}}{z+2},\quad m\geq 0,n>0,\\
\mathbf{s}^{\text{\tiny 0n1}}_{\text{\tiny 1}}&=0, \quad n\geq 0, \\
\mathbf{s}^{\text{\tiny mn1}}_{\text{\tiny 1}}&=\frac{z\left( (z^{2}+1) \mathbf{b}^{\text{\tiny m(n+1)1}}_{\text{\tiny L}} -2 \mathbf{b}^{\text{\tiny (m-1)(n+1)1}}_{\text{\tiny T}}\right)}{(z+2)(z-1)},\quad m>0,n\geq 0,\\
\mathbf{s}^{\text{\tiny mn2}}_{\text{\tiny 1}}&=-\frac{2(2z-1)(z^{2}+1)\mathbf{b}^{\text{\tiny (m+1)n1}}_{\text{\tiny L}}}{3z(z-1)},\quad m\geq 0,n \geq 0,\\
\mathbf{s}^{\text{\tiny mn1}}_{\text{\tiny 2}}&=-\frac{2(z^{2}+1)\mathbf{b}^{\text{\tiny (m+1)n1}}_{\text{\tiny L}}}{(z+2)(z-1)},\quad m\geq 0,n\geq 0,\\
\mathbf{s}^{\text{\tiny 0n2}}_{\text{\tiny 2}}&=0, \quad n\geq 0,\\
\mathbf{s}^{\text{\tiny mn2}}_{\text{\tiny 2}}&=-\frac{z\left( (z^{2}+1) \mathbf{b}^{\text{\tiny m(n+1)1}}_{\text{\tiny L}} -\mathbf{b}^{\text{\tiny (m-1)(n+1)1}}_{\text{\tiny T}}\right)}{(z+1)(z-4)}, \quad m>0,n \geq 0.
\end{align*}
Thus, we find that the scalar modes are parametrised by two coefficients $\mathbf{b}^{\text{\tiny mn1}}_{\text{\tiny L}}$ for $m>0,n\geq0$ and $\mathbf{b}^{\text{\tiny mn1}}_{\text{\tiny T}}$ for $m\geq 0,n>0$. The solution  $\mathbf{s}^{\text{\tiny mn2}}_{\text{\tiny 2}}$ has a singularity at $z=4$, but that do not ''leak'' into the conserved quantities computed below, so we avoid going over the same analysis for $z=4$. Moreover, at $z\geq 4$ that mode is already constant or growing so it should be discarded. The modes $\mathbf{s}^{\text{\tiny mn1}}_{\text{\tiny 3}}$ and $\mathbf{s}^{\text{\tiny mn1}}_{\text{\tiny 4}}$ do not contribute to the stress energy tensor. As a matter of fact, these correspond to the pure gauge components that are discussed in constant pertubations so they also seem to stay as pure gauge in this case.

Plugging the solutions \eqref{solbL}-\eqref{sols4} in \eqref{E1comp}-\eqref{PiScomp} yield the following results for the components of the stress energy tensor
\begin{align*}
\mathcal{E}^{(0,n)}_{1}&=0,\quad n\geq 0,\\
 \mathcal{E}^{(m,n)}_{1}&=4(z^{2}+1)\mathbf{b}^{\text{\tiny mn1}}_{\text{\tiny L}},\quad m>0,\,n \geq 0,\\
 \mathcal{E}^{(m,n)}_{2}&=0,\quad m\geq 0,n\geq 0,\\
\mathcal{E}^{x(m,n)}&=-4(z^{2}+1)\mathbf{b}^{\text{\tiny (m+1)n1}}_{\text{\tiny L}},\quad m\geq 0,\,n\geq 0,\\
\mathcal{P}_{x}^{(0,n)}&=0,\quad n\geq 0,\\
\mathcal{P}_{x}^{(m,n)}&=-2z\left[(z^{2}+1)\mathbf{b}^{\text{\tiny m(n+1)1}}_{\text{\tiny L}}-2 \mathbf{b}^{\text{\tiny (m-1)(n+1)1}}_{\text{\tiny T}}\right],\quad m>0,\,n\geq 0,\\
\Pi^{(0,n)}_{L}&=0,\quad n\geq 0,\\
\Pi^{(m,n)}_{L}&=2z(z^{2}+1)\mathbf{b}^{\text{\tiny mn1}}_{\text{\tiny L}}, \quad m>0,\, n\geq 0,\\
\Pi^{(m,0)}_{T}&=-2z(z^{2}+1)\mathbf{b}^{\text{\tiny (m+1)01}}_{\text{\tiny L}},\quad m\geq 0,\\
\Pi^{(m,n)}_{T}&=-4z \mathbf{b}^{\text{\tiny mn1}}_{\text{\tiny T}},\quad m\geq 0,\,n>0,\\
\Pi^{(m,n)}_{S}&=0, \quad m\geq,\, n\geq 0,
\end{align*}
which are all evaluated at infinity. It is now straightforward to verify that the conservation \eqref{finconsv1}, 
\eqref{finconsv2} and scaling \eqref{finconsv3} equations  are satisfied. 

\subsubsection{Vector modes}
The next and final step in general perturbations is the vector mode solutions. The relevant components of the energy momentum complex read
\begin{align}
\mathcal{E}^{y}&=-k\omega r^{z+2}\left[r^{2z-1}\mathsf{v}_{1}'-r\mathsf{v}_{3}'
-\frac{2(z-1)}{\sqrt{z+1}}\left(\frac{ \mathsf{v}_{6}}{r}+r^{z-2}\mathsf{v}_{4}\right)\right]e^{i(\omega t+k x)},\label{eny}\\
\mathcal{P}_{y}&=-k \omega r^{z+2}\left[r \mathsf{v}_{1}'-(2z-1)r^{3-2z}\mathsf{v}_{3}'-2(z-1)r^{1-2z}\mathsf{v}_{8}\right]e^{i(\omega t+k x)},\label{momy}\\
\Pi^{x}{}_{y}&=- r^{z+2}\left[zr\mathsf{v}_{2}'+2(z-1)\left(r\mathsf{v}_{5}'+2 \mathsf{v}_{5}+\sqrt{z+1}\mathsf{v}_{7}+\frac{k^{2}\mathsf{v}_{8}}{r}\right)\right]e^{i(\omega t+k x)}.\label{spay}
\end{align}
This time, the only conservation law needed to be fullfilled is the conservation of momenta along $y$ direction
\begin{align}
\omega\mathcal{P}_{y}+k \Pi^{x}{}_{y}=0,
\end{align}
which can be written as a combination of $\mathcal{E}^{2}_{r}$ and $\mathcal{E}_{x_{2}r}$. On the other hand, $\mathcal{E}^{y}$ is not constrained. We take the expansions of \eqref{eny}-\eqref{spay} in the form
\begin{align}
\mathcal{E}^{y}={} &\sum_{m,n}k^{2m}\omega^{2n}\mathcal{E}^{y(m,n)},\\
\mathcal{P}_{y}={}&\sum_{m,n}k^{2m+1}\omega^{2n+1}\mathcal{P}_{y}^{(m,n)},\\
\Pi^{x}{}_{y}={}&\sum_{m,n}k^{2m}\omega^{2n}\Pi^{x(m,n)}_{y},
\end{align}
which amount to the conservation equation
\begin{align}
\mathcal{P}_{y}^{(m,n)}+\Pi^{x(m,n+1)}_{y}=0. \label{consvy}
\end{align}

The structure of expanded field equations remains similar so, following the same strategy as before we write down the contributing solutions 
\begin{align}
\mathsf{v}_{1}^{(m,n)}(r)&=\mathbf{v}^{\text{\tiny mn1}}_{\text{1}}\,r^{-(z+2)}
+\mathbf{v}^{\text{\tiny mn2}}_{\text{1}}\,r^{-3z}+\mathcal{O}(r^{p}),\\
\mathsf{v}_{2}^{(m,n)}(r)&=\mathbf{v}^{\text{\tiny mn1}}_{\text{2}}\,r^{-(z+2)}+\mathcal{O}(r^{p}),\\
\mathsf{v}_{3}^{(m,n)}(r)&=\mathbf{v}^{\text{\tiny mn1}}_{\text{3}}\,r^{-(z+2)}
+\mathbf{v}^{\text{\tiny mn2}}_{\text{3}}\,r^{-(4-z)}+\mathcal{O}(r^{p}),\\
\mathsf{v}_{4}^{(m,n)}(r)&=\mathbf{v}^{\text{\tiny mn1}}_{\text{4}}\,r^{-2z}+\mathcal{O}(r^{p}),\\
\mathsf{v}_{5}^{(m,n)}(r)&=\mathbf{v}^{\text{\tiny mn1}}_{\text{5}}\,r^{-(z+2)}+\mathcal{O}(r^{p}),\\
\mathsf{v}_{6}^{(m,n)}(r)&=\mathbf{v}^{\text{\tiny mn1}}_{\text{6}}\,r^{-(z+1)}+\mathcal{O}(r^{p}),\\
\mathsf{v}_{7}^{(m,n)}(r)&=\mathbf{v}^{\text{\tiny mn1}}_{\text{7}}\,r^{-(z+2)}+\mathcal{O}(r^{p})\\
\mathsf{v}_{8}^{(m,n)}(r)&=\mathbf{v}^{\text{\tiny mn1}}_{\text{8}}\,r^{-(z+1)}
+\mathbf{v}^{\text{\tiny mn2}}_{\text{8}}\,r^{-(3-z)}+\mathcal{O}(r^{p}).
\end{align}
In addition to the $\mathbf{v}^{\text{\tiny mn2}}_{\text{3}}$ which is growing for $z\geq 4$, we need to keep an eye on $\mathbf{v}^{\text{\tiny mn2}}_{\text{8}}$ mode for $z\geq 3$. Feeding these into field equations and grinding through several orders we arrive at
\begin{align*}
\mathbf{v}^{\text{\tiny 0n1}}_{\text{1}}&=\mathbf{v}^{\text{\tiny m02}}_{\text{1}}=0,\quad m\geq 0,\, n\geq 0,\\
\mathbf{v}^{\text{\tiny mn2}}_{\text{1}}&=-\frac{z(z^{2}+z+1)(2z^{2}+9z+13)}{(z+2)(2z^{4}+2z^{3}-z^{2}+3z-2)}\mathbf{v}^{\text{\tiny (m+1)(n-1)1}}_{\text{1}},\quad m\geq 0,\, n>0,\\
\mathbf{v}^{\text{\tiny m01}}_{\text{2}}&=\mathbf{v}^{\text{\tiny 0n1}}_{\text{2}}=0,\quad m\geq 0,\, n\geq 0,\\
\mathbf{v}^{\text{\tiny mn1}}_{\text{2}}&=-\frac{z(z-1)(4z+11)(z+1)^{2}}{(z+2)(2z^{4}+2z^{3}-z^{2}+3z-2)}
\mathbf{v}^{\text{\tiny m(n-1)1}}_{\text{1}},\quad m>0,\, n>0,\\
\mathbf{v}^{\text{\tiny m01}}_{\text{3}}&=\mathbf{v}^{\text{\tiny 0n2}}_{\text{3}}=0,\quad m\geq 0,\, n\geq 0,\\
\mathbf{v}^{\text{\tiny mn1}}_{\text{3}}&=-\frac{(z^{2}+z+1)(4z^{2}+7z-3)}{(2z^{4}+2z^{3}-z^{2}+3z-2)}
\mathbf{v}^{\text{\tiny (m+1)(n-1)1}}_{\text{1}},\quad m\geq 0,\, n> 0,\\
\mathbf{v}^{\text{\tiny mn2}}_{\text{3}}&=-\frac{z+2}{z-4}\mathbf{v}^{\text{\tiny mn1}}_{\text{1}},\quad m>0,\, n\geq 0,\\
\mathbf{v}^{\text{\tiny mn1}}_{\text{4}}&=0,\quad m\geq 0,\quad n\geq 0,\\
\mathbf{v}^{\text{\tiny m01}}_{\text{5}}&=\mathbf{v}^{\text{\tiny 0n1}}_{\text{5}}=0,\quad m\geq 0,\, n\geq 0,\\
\mathbf{v}^{\text{\tiny mn1}}_{\text{5}}&=-\frac{3(z^{4}+8z^{3}+17z^{2}+12z-2)}{2(z+2)(2z^{4}+2z^{3}-z^{2}+3z-2)}
\mathbf{v}^{\text{\tiny m(n-1)1}}_{\text{1}},\quad m>0,\, n>0,\\
\mathbf{v}^{\text{\tiny m01}}_{\text{6}}&=0,\quad m \geq 0,\\
\mathbf{v}^{\text{\tiny mn1}}_{\text{6}}&=\frac{2\sqrt{z+1}(z^{2}+z+1)(2z^{3}+9z^{2}+10z-3)}
{(z+2)(2z^{4}+2z^{3}-z^{2}+3z+2)}\mathbf{v}^{\text{\tiny (m+1)(n-1)1}}_{\text{1}},\quad m\geq 0,\,n>0,\\
\mathbf{v}^{\text{\tiny m01}}_{\text{7}}&=\mathbf{v}^{\text{\tiny 0n1}}_{\text{7}}=0,\quad m\geq 0,\, n\geq 0,\\
\mathbf{v}^{\text{\tiny mn1}}_{\text{7}}&=\frac{(z+1)^{3/2}(z^{2}+z+1)}{(2z^{4}+2z^{3}-z^{2}+3z-2)}\mathbf{v}^{\text{\tiny m(n-1)1}}_{\text{1}},\quad m > 0,\,n>0,\\
\mathbf{v}^{\text{\tiny m01}}_{\text{8}}&=0,\quad m\geq 0,\\
\mathbf{v}^{\text{\tiny mn1}}_{\text{8}}&=-\frac{2(z+1)(z^{2}+z+1)(2z^{3}+9z^{2}+10z-3)}{(2z^{5}+6z^{4}+3z^{3}+z^{2}+4z-4)}\mathbf{v}^{\text{\tiny (m+1)(n-1)1}}_{\text{1}},\quad m\geq 0,\, n>0,\\
\mathbf{v}^{\text{\tiny mn2}}_{\text{8}}&=0,\quad m\geq 0,\,n \geq 0.
\end{align*}
In this case the solutions are parametrized by only $\mathbf{v}^{\text{\tiny mn1}}_{\text{1}}$ for $m>0,\,n\geq 0$. Similar to the constant perturbations, the mode $\mathbf{v}^{\text{\tiny mn1}}_{\text{4}}$ corresponding to the electric charge does not contribute here. Moreover, the mode $\mathbf{v}^{\text{\tiny mn2}}_{\text{8}}$ which we mentioned before has no nontrivial solution. Gathering all the solutions, we compute the vector components of stress tensor 
\begin{align}
\mathcal{P}_{y}^{(0,n)}&=0,\quad n\geq 0,\\
\mathcal{P}_{y}^{(m,n)}&=-2(z+2)(z-1)\mathbf{v}^{\text{\tiny mn1}}_{\text{1}},\quad m>0,\, n\geq 0,\\
\Pi^{x}{}_{y}^{(m,0)}&=\Pi^{x}{}_{y}^{(0,n)}=0,\quad m\geq 0,\,\,n\geq 0,\\
\Pi^{x}{}_{y}^{(m,n)}&=2(z+2)(z-1)\mathbf{v}^{\text{\tiny m(n-1)1}}_{\text{1}},\quad m>0,\, n>0,\\
\mathcal{E}^{y(m,0)}&=0,\quad m\geq 0,\\
\mathcal{E}^{y(m,n)}&=\frac{6z(z-1)(z+3)(z^{2}+z+1)}{2z^{4}+2z^{3}-z^{2}+3z-2}
\mathbf{v}^{\text{\tiny (m+1)(n-1)1}}_{\text{1}},\quad m \geq 0,\, n> 0.
\end{align}
It is now evident that \eqref{consvy} is satisfied.

In summary, the preceding two sections showed that the conservation laws are satisfied on-shell. We perform this by first verifying that the conservation equations are a combination of full field equations. Then, we solve the expanded field equations for the modes that contribute to the stress-energy tensor and compute the values of components explicitly. It is worth emphasizing that we argued only from the structure of the expanded field equations that the particular solutions would be suppressed. We did not solve the equations explicitly since the number of terms and complications of the equations proliferate as we go on higher orders in $(n,m)$. It is possible (especially for the vector modes) that some of the particular solutions may grow large and spoil the behavior of stress-energy tensor at infinity. In any case, as demonstrated in \cite{Ross:2009ar}, it is possible to kill these divergences by adding $S_{deriv}$ to the action and adjust the free parameters accordingly. 
\endgroup
\section{Numerical black holes and thermodynamics}{\label{numblackhole}}
The final task we undertake will be to study the conserved quantities and thermodynamics of the numerical black hole solutions found in \cite{Devecioglu:2014iia}. The solutions admit Lifshitz spacetime as a background and for a fixed horizon radius, depend on one parameter, the gauge field strength at the horizon. The behavior of solutions differs among horizon topologies; that being said, in this work we will restrict our attention to the planar ones, as the stress-energy complex is constructed for the planar boundary. In \cite{Fan:2015yza}, it was discussed (through linear analysis) that unlike their AdS counterparts in other models \cite{Fan:2014ixa} which can have non-zero YM charges, the Lifshitz black holes are analogous to the Minkowski family \cite{Bizon:1990sr}, i.e., they do not possess a global YM charge. In this section, our first aim is to compute the numerical value of the energy of the solutions and show they are finite, obeying scale relations. Then, employing the energies for different horizon radius, we will demonstrate that the alternative Smarr relation and the first law of thermodynamics hold without any YM hair modification, extending the result of \cite{Fan:2015yza} into the non-linear regime. 

Let us start with the redefinitions of metric and gauge field functions which will help us to cast them in more convenient forms for numerical purposes. The metric is chosen in the following way
\begin{align}
 ds^{2}=-r^{2z}f(r)^{2}dt^{2}+r^{2}(d x^{2}+d y^{2})+\dfrac{g(r)^{2}dr^{2}}{r^{2}},\label{bhmetric}
\end{align}
and the gauge field \eqref{planarsymgauge} redefinitions are as follows
\begin{align}
R(r)=\sigma r h(r),\quad R(r)^{\prime}=\sigma j(r).\label{bhgauge}
\end{align}
The function $j(r)$ is defined to reduce the boundary value problem to an initial value problem and employ the shooting method for numerical solution of field equations. Thus, as $r\rightarrow\infty$, all functions will asymptote to unity to have Lifshitz background. The first order field equations for these choices read\footnote{The reader is referred to \cite{Devecioglu:2014iia} for details of the numerical solutions.}
\begin{align}
r f(r)^{\prime} & =-f(r)\Big((z-1)-\dfrac{j(r)^{2}}{2}(z-1)+\dfrac{g(r)^{2}h(r)^{4}}{4}(z^{2}-1)-\dfrac{g(r)^{2}}{4}(3+2z+z^{2})
+\dfrac{3}{2}\Big),\label{fulleqns1}\\
r j(r)^{\prime} & =  j(r)+g(r)^{2}h(r)^{3}(z+1)-
\dfrac{g(r)^{2}j(r)}{2}(z^{2}+2z+3)+\dfrac{g(r)^{2}h(r)^{4}j(r)}{2}(z^{2}-1),\label{fulleqns2}\\
r g(r)^{\prime} & =  \dfrac{g(r)j(r)^{2}}{2}(z-1)
+\dfrac{g(r)^{3}h(r)^{4}}{4}(z^{2}-1)-g(r)^{3}(3+2z+z^2)+\dfrac{3 g(r)}{2},\label{fulleqns3}\\
rh(r)^{\prime}&=j(r)-h(r).\label{fulleqns4}
\end{align}
Here, the equation \eqref{fulleqns1} is linear in function $f(r)$ so the overall normalization of $f$ is not fixed. Thus, the numerical value of $f(r)\rightarrow f_{\infty}$ will be normalized to unity at infinity by dividing $f_{\infty}$, which corresponds to rescaling of the time coordinate. In numerical integration, we will set the initial value at the horizon, i.e., $f(R_0)=1$ then deal with the normalization at the end.

Before moving on to numerical solutions, let us shortly mention about the horizon expansions which will shed light on the initial values of functions and shooting parameter. Provided $g_{tt}$ and $g_{rr}$ components of \eqref{bhmetric} have a simple zero and a simple pole, the black hole will be non-extremal. This argument leads to the following expansions at the horizon 
\begin{align}
f(r)&=\sqrt{r-R_{0}}\,\sum_{n=0}^{\infty}f_{n}(r-R_{0})^{n},\label{fexpand}\\
g(r)&=\dfrac{1}{\sqrt{r-R_{0}}}\,\sum_{n=0}^{\infty}g_{n}(r-R_{0})^{n}.\label{gexpand}
\end{align}
Incorporating these into field equations, one can study the series solutions \cite{Devecioglu:2014iia} and extract info on initial values.

\begin{figure}
\centering
\includegraphics[scale=1.5]{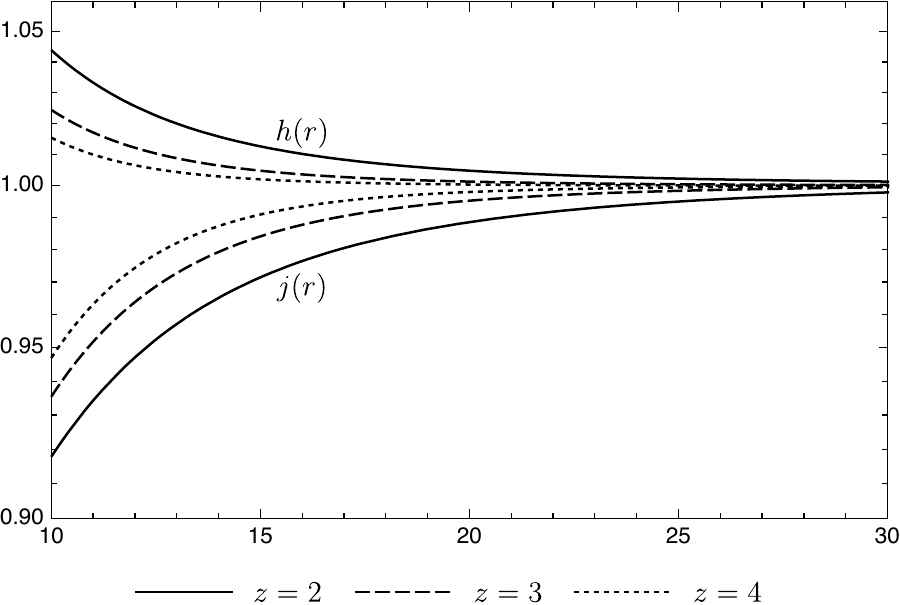}
\caption{The figure plots the gauge field functions $h(r)$ and $j(r)$ as a function of radial distance $r$. This is an example of a large black hole with $R_{0}=10$ for different values of $z$.}\label{fig1}
\end{figure}

\begin{figure}
\center
\includegraphics[scale=1.5]{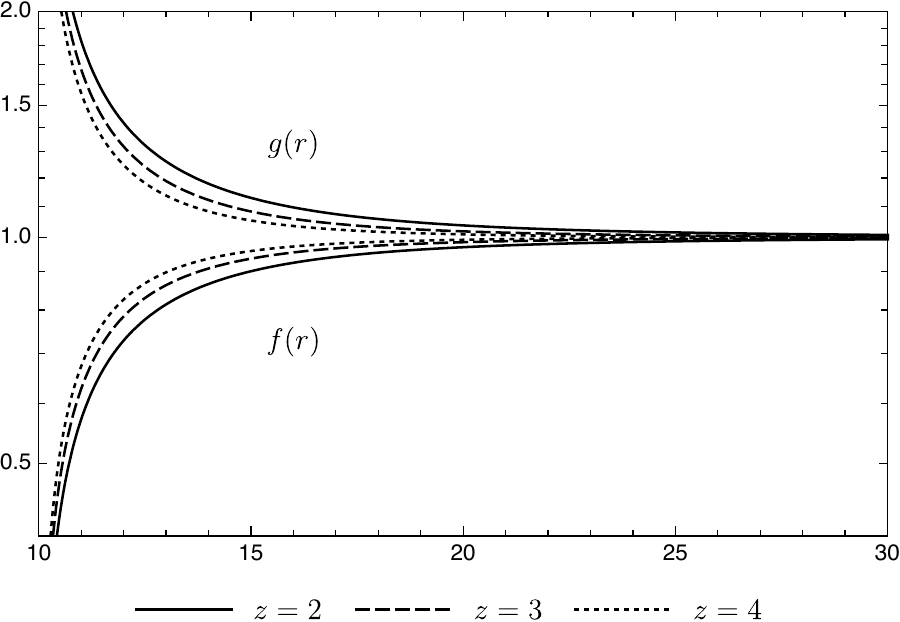}
\caption{The figure shows the metric functions $f(r)$ and $g(r)$ as a function of radius $r$ with $R_{0}=10$.}\label{fig2}
\end{figure}

The next step is to numerically integrate the system of equations \eqref{fulleqns1}-\eqref{fulleqns4} by fixing one of the free parameters, i.e. the event horizon radius $R_{0}$ and make the functions $g,h,j$ converge to unity asymptotically, by fine-tuning the initial value of $h(R_{0})=h_{0}$. Setting $R_{0}=10$ we find solutions for $z=2,3,4$ and represented them in Figures \ref{fig1},\ref{fig2}. The value of the shooting parameter $h_{0}$ varies with different $z$ but stays the same for large and small planar black holes. On the other hand, for other topologies, the shooting parameter and the behavior of functions vary for small and large black holes with respect to the length scale $L$ (which should appear in \eqref{bhmetric} but fixed to $L=1$ in our discussion).

\begin{figure}
\center
\includegraphics[scale=1.5]{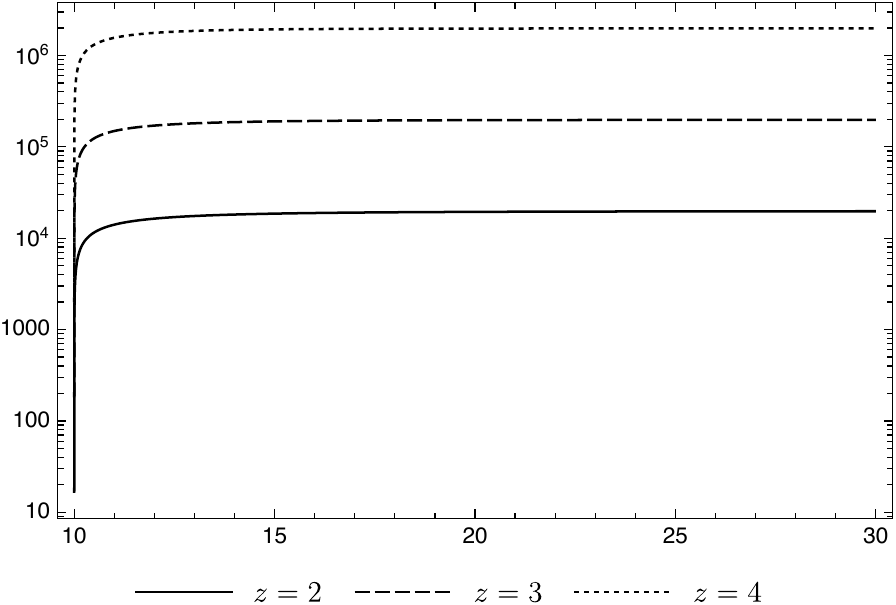}
\caption{The energy of a black holes with $R_{0}=10$. Figure shows the energy for different cases with $z=2,3,4$.}\label{fig3}
\end{figure}

\begin{figure}
\center
\includegraphics[scale=1.5]{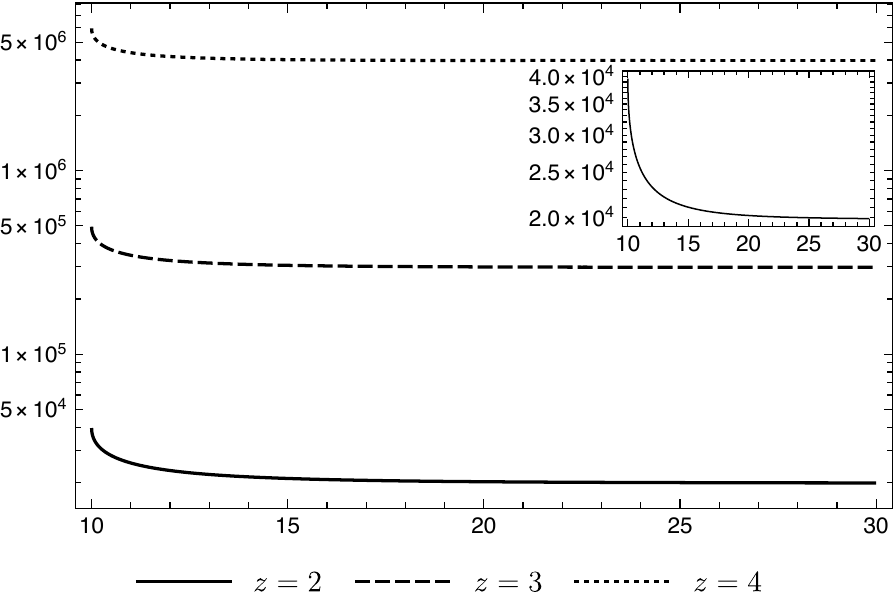}
\caption{The figure illustrates the $\Pi^{x}{}_{x}=\Pi^{y}{}_{y}$ components of the black holes with $R_{0}=10$. The smaller figure is a close up to demonstrate the behaviour for $z=2$. }\label{fig4}
\end{figure}

Having solutions in hand, we now appeal to the stress tensor definitions \eqref{firstTset}, \eqref{secTset}  and compute the non-zero components
\begin{align}
E &= r^{z+2}f  \left(z+3-(z-1) h^4-\frac{4}{g}\right)\label{energybh},\\
\Pi^{x}{}_{x} &=\Pi^{y}{}_{y} =\frac{r^{z+2}}{g} \left[2 r f'+f \left(g \left((z-1) h^4-z-3\right)
+2 \left(z+1-r (z-1) h h'-(z-1) h^2\right)\right)\right]\label{spatialbh}.
\end{align}
As in the case of field equation \eqref{fulleqns1} the energy and diagonal components of the spatial stress tensor are linear in $f$ and will be normalised with $f_{\infty}$ of the corresponding $z$ value. After evaluating the numerical solutions on \eqref{energybh} and \eqref{spatialbh}, we plot the results in Figures \ref{fig3}, \ref{fig4}. 

First relation we check is whether the presence of a black hole breaks the scale invariance or not. It is evident from the values in the Table \ref{table:1} that the scaling Ward identity $z\mathcal{E}=\Pi^{i}{}_{i}$ still holds within the bounds of numerical errors.
\begin{table}[h!]
\centering
\begin{tabularx}{0.7\textwidth} {| >{\centering\arraybackslash}X  | >{\centering\arraybackslash}X | >{\centering\arraybackslash}X | >{\centering\arraybackslash}X | }
   \hline
   & $z=2$ & $z=3$ & $z=4$ \\
   \hline
  $z\mathcal{E}$ & $3.940 \times 10^{4}$ & $5.923 \times 10^{5}$ & $7.921 \times 10^{6}$ \\
   \hline
  $\Pi^{i}{}_{i}$  & $3.952 \times 10^{4}$ & $5.927 \times 10^{5}$  & $7.923 \times 10^{6}$ \\
    \hline
\end{tabularx}
 \caption{The numerical values of $z\mathcal{E}$ and $\Pi^{i}{}_{i}$ at $r \rightarrow\infty$.}
 \label{table:1}
\end{table}

\begin{figure}
\center
\includegraphics[scale=1.5]{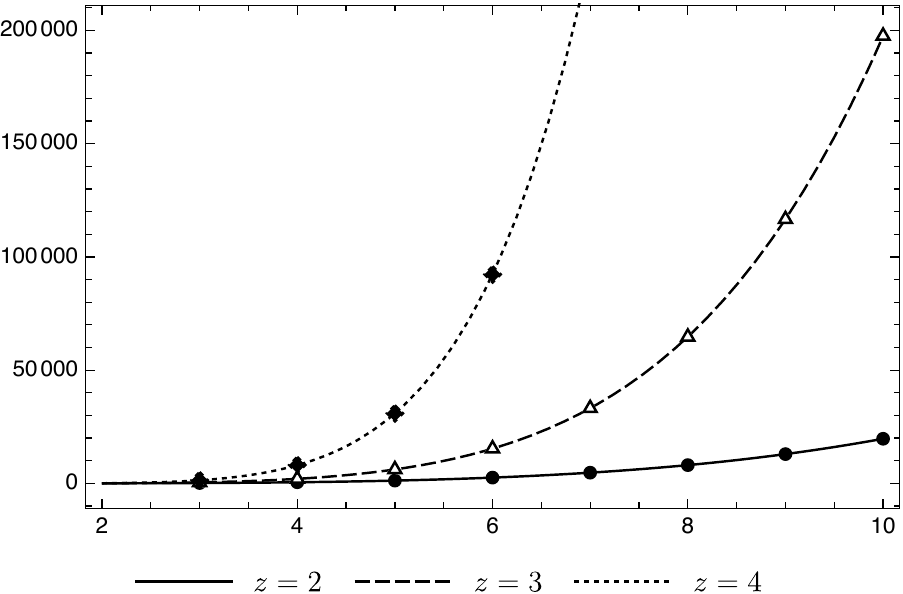}
\caption{The circle, square, and triangle points are numerical values generated from \eqref{energybh} for different horizon values, while solid, dashed, dotted curves are quartic, quintic and sextic polynomials, respectively. As demonstrated, the numerical values fit nicely to their corresponding polynomial curves.}\label{fig5}
\end{figure}

As a second and final check, we verify the thermodynamic relations of these solutions. The energy $\mathcal{E}$ defined here is shown to agree with the thermodynamic energy density obtained by the Euclidean version of the black holes \cite{Ross:2009ar}. The proof is quite general so, it is possible to extend and apply it to our case. For the planar Lifshitz black holes, i.e. $k=0$, it was also shown that \cite{Brenna:2015pqa}  the thermodynamic quantities are consistent with an alternative Smarr relation
\begin{align}
(D+z-2)E=(D-2)TS,\label{smarr}
\end{align}
and an accompanying first law
\begin{align}
dE=T dS.\label{flaw}
\end{align}
These two equalities have the extended versions when the solutions admit global charges, but as we discussed before, in \cite{Fan:2015yza} by studying the linearized solutions, it was argued that the exclusion of the $\fc_{3}$ mode (as it diverges for $r\rightarrow\infty$) implies the lack of global YM charge. In accordance with this reasoning \eqref{smarr}, \eqref{flaw} should hold in their simplest form. In order to check this claim, we first compute the Hawking temperature through Wick rotation with an assumption of regularity at the horizon 
\begin{align}
T=\frac{r^{z+1}}{4\pi}\sqrt{f(r)^{\prime}g(r)^{\prime}}\Bigg |_{r=R_{0}},
\end{align}
and the entropy is $S=\pi R_{0}^{2}$ where we suppose suitable identifications and volume renormalizations. From near horizon expansions \eqref{fexpand}, \eqref{gexpand} it is easy to see $T\sim R_{0}^{z}$. Then the first law \eqref{flaw} implies the energy should scale as $E\sim R_{0}^{z+2}$, which is in agreement with the $L^{D+z-2}$ scaling that can be obtained from the Euler's relation in \eqref{smarr}. This is different from $L^{D-3}$ scaling in AdS as alluded in \cite{Brenna:2015pqa}. On the numerical side, we verify this feature by plotting the energy of the black holes with respect to their horizon radius having different dynamical exponents in Fig. \ref{fig5}. 

It remains to verify the Smarr relation; to that end, we opted to compute the left and right-hand side of \eqref{smarr} numerically and represent them in the Table \ref{table:2}. For the sake of clarity, we also calculate the ratio $2TS/(z+2)E$ (which is not unity, as we did not include the physical constants in our numerical computations) to show that the Smarr relation holds \eqref{smarr}.
\begin{table}[h!]
\centering
\begin{tabularx}{0.7\textwidth} {| >{\centering\arraybackslash}X  | >{\centering\arraybackslash}X | >{\centering\arraybackslash}X | >{\centering\arraybackslash}X | }
   \hline
   & $z=2$ & $z=3$ & $z=4$ \\
   \hline
  $(z+2)\mathcal{E}$ & $7.425 \times 10^{8}$ & $9.188 \times 10^{9}$ & $1.105 \times 10^{11}$ \\
   \hline
  $2TS$  & $9.296 \times 10^{13}$ & $1.149 \times 10^{15}$  & $1.381 \times 10^{16}$ \\
    \hline
     $2TS/(z+2)E$  & $1.251 \times 10^{5}$ & $1.250 \times 10^{5}$  & $1.250 \times 10^{5}$ \\
    \hline
\end{tabularx}
 \caption{The numerical values of $(z+2)\mathcal{E}$ and $2TS$ for $z=2,3,4$. The final row of the table contains the ratio $2TS/(z+2)E$. Here we did not divide the first two rows with $f_{\infty}$ as their ratio will be the same.}
 \label{table:2}
\end{table}

With this final result, we demonstrated that the thermodynamic relations hold without an additional YM charge. As advertised earlier in \cite{Fan:2015yza}, this makes the numerical black holes of \cite{Devecioglu:2014iia} hairy solutions. The stability of these black holes is another important question that can save the no-hair theorem.

In short, the energy definition is shown to work well in the non-linear regime, producing finite results with the scaling properties left intact. Moreover, the thermodynamical energy is in agreement with the Smarr formula and the first law, confirming the $SU(2)$ hair of the black holes.

\section{Conclusions}{\label{conclusions}}
In this work, we have computed the holographic stress-energy tensor of the EYM model with Lifshitz backgrounds as a solution. After setting the stage for a properly defined tensor, the main results can be outlined as follows.

First, we study the linearised field equations to show that the action and the stress-energy tensor stay finite for a more general class of asymptotically Lifshitz spacetimes. The choice for the form of perturbation plays an essential role in decomposing field equations into scalar, vector, and tensor modes. For constant perturbations, we were able to solve the linearised equations in their full forms and choose the modes that asymptote to the Lifshitz backgrounds. We can summarise the modes as:
\begin{itemize}
\item The modes $\fc_{1},\ft_{d1},\ft_{o1}$ have fall rate $(z+2)$ and contribute to the $\mathcal{E},\Pi^{i}{}_{j}$ components of the stress tensor.  
\item The mode $\fc_{1i}$ have the fall rates $(z+2)$, $(4-z)$ and contribute to $\mathcal{P}_{i}$.
\item The mode $\fc_{2i}$ have the fall rates $(z+2)$, $3z$ and contribute to $\mathcal{E}^{i}$.  
\item The mode $\fc_{2}$  has a fall rate $(z+2+\beta_{z})$ for $z>1$, and the modes $\ft_{d2},\ft_{o2}$ decay as 
$(z+2+\xi_{z})$ for $z\geq 2(3+2\sqrt{2})$. These do not contribute to the stress tensor and related to the expectation value of the operator associated with the $SU(2)$ gauge field.
\end{itemize}
On the other hand, in general perturbations, the complicated structure does not allow for an exact solution. Nevertheless, it was possible to solve the system through an expansion in $k$ and $\omega$. In both instances (constant and generalised), the conservation and scaling laws are shown to be satisfied on-shell. 

In the second part of the work, we turned our attention to the numerical black hole solutions of \cite{Devecioglu:2014iia}. Focusing on the planar ones, we find the energy and the spatial stress tensor for different $z$ values and show that they are finite while satisfying the scaling identity $z\mathcal{E}=\Pi^{i}{}_{i}$. The results prove that the stress tensor definition is reliable in the non-linear regime. A more non-trivial check was the thermodynamic equalities satisfied by these black holes. The Smarr relation and the first law hold without the YM parameter so that the solutions are hairy, like their asymptotically flat and AdS counterparts. 

There are several directions for further exploration in this model. First, on the holography side, using the results of the mode analysis as a starting point, it is now possible to look for the divergences in the one-point functions and perform holographic renormalisation. It will also be interesting to see the structure of TNC geometry that will be conjured on the boundary. Along these lines, it can be helpful to investigate the eleven-dimensional model \cite{Pope:1985bu}, which might help to extract information about the sources and VeVs.

Finally, hairy black holes of AdS are used in modelling the second-order transitions \cite{Gubser:2008px}, where an abelian gauge symmetry is spontaneously broken near a black hole horizon in ADS using a condensate of $SU(2)$ gauge fields. A similar setup where the AdS solution is replaced with a hairy Lifshitz black hole can now be considered by turning on the electric part $Q(r)$ in the gauge field ansatz. We plan to return these issues in the future. 

\begin{acknowledgments}
 I am grateful to Simon F. Ross for the detailed explanation of his work, comments and suggestions. I also thank G{\"o}khan Alka\c{c}, {\"O}zg{\"u}r Sar{\i}o\u{g}lu for discussions. Finally I thank G{\"o}khan Alka\c{c} for his critical reading of the manuscript. This work was supported by the National Natural Science Foundation of China under Grant No. 11875136 and the Major Program of the National Natural Science Foundation of China under Grant No. 11690021.
\end{acknowledgments}
\newpage
  
\appendix
\section{Linearized quantities}{\label{appendix1}}
\begingroup
\allowdisplaybreaks
Here we present some of the linearized objects that are used to compute the perturbative action and the field equations.
\begin{align}
K=\bar{K}+K^{L}=&(z+2)+\frac{r}{2}\partial_{r}[\hat{h}_{tt}+\hat{h}^{i}{}_{i}],\\
R=\bar{R}+R^{L}=&-2(3+z(2+z))-r\left((3+2z)\partial_{r}\hat{h}_{tt}+(4+z)\partial_{r}\hat{h}^{i}{}_{i}+r\partial_{r}^{2}[\hat{h}_{tt}+\hat{h}^{i}{}_{i}]\right),\\
F^{\Lambda}_{\mu\nu}F^{\mu\nu}_{\Lambda}=&\bar{F}^{\Lambda}_{\mu\nu}\bar{F}^{\mu\nu}_{\Lambda}+(F^{\Lambda}_{\mu\nu}F^{\mu\nu}_{\Lambda})^{L}=2(z+1)\left(
3+z+2(2+z)\hat{b}^{i}{}_{i}+r\partial_{r}[\hat{h}^{i}{}_{i}+2\hat{b}^{i}{}_{i}]\right),\\
F^{\Lambda}_{ab}F^{ab}_{\Lambda}=&4(z+1)^{2}(1+2\hat{b}^{i}{}_{i}).
\end{align}
\section{Field equations for generalised perturbations}{\label{appendix2}}
\subsection{Field equations for the scalar modes}
In this section we provide the scalar mode field equations for the generalised perturbations. First, let us call the the linearized equations for the gravity sector as $\mathcal{E}_{\mu\nu}$ and for the matter sector $\mathcal{E}_{\mu}^{\Lambda}$. The field equations for the scalar modes read
\begin{flalign}
    r^{2} \left(f''+k_{L}''\right)+2 (z+1) r f'+(z+3) r k_{L}'-4 \left(z^2-1\right)b_{L}  ={} \frac{ k^2  }{ 2r^2 }f-\frac{2k^{2}(z^{2}-1)s_{3}}{r\sqrt{z+1}}\nonumber\\
          -\frac{\omega^{2} }{ r^{2 z}}k_{L}-\frac{ k^2 \omega^2  }{r^2}s_{1}\label{eq:1}\\
    r^{2} (2b_{L}''+ k_{L}'')+ r(z+3)(2b_{L}'+k_{L}')+r f'-4 (z+1)b_{L} ={} -\frac{k^{2}}{r}\left(rs_{4}^{\prime}+zs_{4}+3\sqrt{z+1}s_{3}\right)\nonumber\\
    +\frac{k^2}{2r^{2}}\left(2b_{L}+k_{L}\right)-\frac{\omega^{2}}{r^{2z}}\left(2b_{L}+k_{L}\right)\nonumber\\
    +\frac{k^{2}\omega^{2}}{r^{2z}}s_{2}-\frac{k^{4}}{r^{2}}(b_{T}+k_{T})\label{eq:2}
\end{flalign}
\begin{flalign}
    2 r k_{L}'+r f'-2 (z-1) b_{L}'+2 (z-1)  b_{L}={}\frac{k^2}{2r^{2}} \left(f+k_{L}\right)
-\frac{ \omega^2 }{r^{2z}}k_{L}\nonumber\\
+\frac{k^{2}(z-1)}{r}(\sqrt{z+1}s_{3}+s_{4})\nonumber\\
-\frac{k^4 }{2r^2}k_{T}+ k^2 \omega^2\left(\frac{s_{2}}{r^{2z}}-\frac{s_1}{r^{2}}\right)\label{eq:3}\\
    r^{2} k_{T}''+4 r (z-1) b_{T}'+ (3z+1)r k_{T}'+4 (z-1) b_{T}-\frac{f}{2r^2}+\frac{2(z-1)s_{4}}{r}={}-\omega^2 \left(\frac{k_{T}}{r^{2z}}+\frac{ s_{1}}{r^2}-\frac{ s_{2}}{r^{2z}}\right)\label{eq:4}\\
    r^{2} (2  b_{T}''+k_{T}'')+(z+1)r k_{T}'+2  (z+3)r b_{T}'+4 (z+1) b_{T}+s_{4}'
    +\frac{zs_{4}}{r}-\frac{\sqrt{z+1}s_{3}}{r}+\frac{ b_{L}}{r^2}+\frac{k_{L}}{2r^2}
   \nonumber\\={}-\frac{\omega^{2}}{r^{2z}}(2b_{T}+k_{T}-s_{2})\nonumber\\
   +\frac{k^{2}}{2r^{2}}(2b_{T}+k_{T})\label{eq:5}\\
    r^{3-z} s_{2}''-r^{1-z}s_{4}'-(z-4) r^{2-z} s_{2}'-r^z s_{1}'+\frac{(z-2)s_{4}}{r^{z}}-\frac{\sqrt{z+1}s_{3}}{r^{z}}
    -\frac{ 1}{2r^{z+1}}(2b_{L}+k_{L})={} \nonumber\\
    k^2 \Big(\frac{b_{T}} {r^{z+1}}+\frac{k_{T} }{2r^{z+1}} +\frac{s_{2}}{r^2}\Big)\label{eq:6}\\
r^{2z} s_{1}''-r^{2}s_{2}''-(z+3)rs_{2}'+r^{2z-1}(3z+1)s_{1}'-\frac{2(z-1)s_{4}}{r}+\frac{2(z^{2}-1)s_{3}}{r\sqrt{z+1}}
    +\frac{k_{L}}{r^{2}}\nonumber\\
    ={} \frac{k^{2}k_{T}}{r^{2}}\label{eq:11}\\
    \frac{2r^{2}(z^{2}-1)s_{3}'}{\sqrt{z+1}}+r(f'+k_{L}'+2(z^{2}-1)s_{4})+(z-1)(2b_{L}+f+k_{L})
    ={} \nonumber\\
    k^{2}(2(z-1)b_{T}+(z-1)k_{T}+rk_{T}')+\omega^{2}(2(z-1)s_{1}+rs_{1}'-r^{3-2z}s_{2}')\label{eq:10}\\
    \frac{2r}{\sqrt{z+1}}(rs_{3}''+(z+3)s_{3}')+2(rs_{4}'+(z+3)s_{3}')={}\frac{k^{2}}{r^{2}}(2rb_{T}+rk_{T}
    +\frac{2s_{3}}{\sqrt{z+1}})
    \nonumber\\
 +\omega^{2}\left(\frac{2s_{1}}{r}-\frac{2s_{2}}{r^{z}}-\frac{2s_{3}}{r^{2z}\sqrt{z+1}}\right)\label{eq:12}
 \end{flalign}

\begin{flalign}
   rk_{L}^{\prime}+2(z-1)b_{L}={}-k^{2}\left(\frac{r^{2z-1}s_{1}^{\prime}}{2}-\frac{1}{2}r^{2}s_{2}'\right)\label{eq:7}\\
   2r^{2}\sqrt{z+1}s_{3}^{\prime}+r(2b_{L}^{\prime}+k_{L}^{\prime})+2(z+1)rs_{4}+2b_{L}+f+k_{L}
   ={}\nonumber\\
   -\frac{k^{2}}{r}(r(2b_{T}-k_{T})+2s_{4}+r^{2}(2b_{T}^{\prime}+k_{T}^{\prime}))\nonumber\\
   +2\omega^{2}(s_{1}+r^{1-2z}(s_{4}+r^{2}s_{2}'))\label{eq:8}\\
   rk_{L}^{\prime\prime}+r(z+3)k_{L}^{\prime}+2(z-1)rb_{L}^{\prime}+2(z^{2}+z-2)b_{L}
   ={}k^{2}\left(\frac{k_{L}}{2r^{2}}+\frac{(z^{2}-1)s_{3}}{r\sqrt{z+1}}-\frac{(z-1)s_{4}}{r}\right)
   -\frac{k^{4}}{2r^{2}}k_{T}
   \label{eq:9}
\end{flalign}
In terms of components \eqref{eq:1}-\eqref{eq:9} corresponds to $(\mathcal{E}_{x_{1}x_{1}}+\mathcal{E}_{x_{2}x_{2}})$, 
$(\mathcal{E}^{1}_{x_{1}}+\mathcal{E}_{x_{2}}^{2})$, $\mathcal{E}_{rr}$, $(\mathcal{E}_{x_{1}x_{1}}-\mathcal{E}_{x_{2}x_{2}})$, $(\mathcal{E}^{1}_{x_{1}}-\mathcal{E}^{2}_{x_{2}})$, $\mathcal{E}^{1}_{t}$, $\mathcal{E}_{tx_{1}}$, $\mathcal{E}_{x_{1}r}$, $\mathcal{E}^{3}_{x_{2}}$, $\mathcal{E}_{tr}$, $\mathcal{E}^{1}_{r}$, $\mathcal{E}_{tt}$ in their order of appearance. Note that, there are twelve equations and nine functions. Therefore three relations constraint this system as follows
\begin{align}
\partial_{r}[\eqref{eq:7}]-\eqref{eq:9}+\frac{z+1}{r}\eqref{eq:7}+\frac{k^{2}}{2}\eqref{eq:11}&=0,\\
r\partial_{r}[\eqref{eq:10}]-(z^{2}-1)r\eqref{eq:12}+\eqref{eq:1}+(z+2)\eqref{eq:10}
+(z-1)\eqref{eq:8}+k^{2}\eqref{eq:4}+\omega^{2}\frac{\eqref{eq:11}}{r^{2z-2}}&=0,\\
r\partial_{r}[\eqref{eq:3}]+(z-1)\eqref{eq:2}-\eqref{eq:1}-z\eqref{eq:9}+(z+2)\eqref{eq:3}-\frac{k^{2}}{2r^{2}}\eqref{eq:10}-\frac{\omega^{2}}{r^{2z+1}}\eqref{eq:7}&=0.
\end{align}
We approach the system of equations by first solving \eqref{eq:10} for $s_{4}$ algebraically. Then, feeding this solution into \eqref{eq:1}-\eqref{eq:12} we expand the equations in $k,\omega$ and solve the homogeneous parts order by order.

\subsection{Field equations for the vector modes}
The equations for the vector modes are as follows
\begin{flalign}
 r^{2}\mathsf{v}_{2}^{\prime\prime}+4(z-1)r\mathsf{v}_{5}^{\prime}+(3z+1)r\mathsf{v}_{2}'+4(z-1)\mathsf{v}_{5}
    ={}-\frac{2k^{2}}{r}(z-1)\mathsf{v}_{7}-\frac{\omega^{2}}{r^{2z}}\mathsf{v}_{2}
    -k^{2}\omega^{2}\left(\frac{\mathsf{v}_{1}}{r^{2}}-\frac{\mathsf{v}_{3}}{r^{2z}}\right)\label{veq:1}\\
     2r^{2}\mathsf{v}_{5}^{\prime\prime}+r^{2}\mathsf{v}_{2}^{\prime\prime}-2\sqrt{z+1}r \mathsf{v}_{8}^{\prime}
    +2(z+3)r\mathsf{v}_{5}^{\prime}+(z+1)r\mathsf{v}_{2}^{\prime}-2\sqrt{z+1}(z+2)\mathsf{v}_{8}+4(z+1)\mathsf{v}_{5}={}\nonumber\\
    k^{2}\left(\frac{\mathsf{v}_{2}}{r^{2}}+\frac{2\mathsf{v}_{5}}{r^{2}}-\frac{2\sqrt{z+1}\mathsf{v}_{6}}{r}\right)
    -\omega^{2}\left(\frac{\mathsf{v}_{2}}{r^{2z}}+\frac{2\sqrt{z+1}\mathsf{v}_{4}}{r^{z}}+\frac{2\mathsf{v}_{5}}{r^{2z}}\right)\label{veq:2}\\
  r^{3-z}\mathsf{v}_{3}^{\prime\prime}-r^{1-z}\mathsf{v}_{7}^{\prime}-(z-4)r^{2-z}\mathsf{v}_{3}^{\prime}-r^{z}\mathsf{v}_{1}^{\prime}+(z-2)r^{-z}\mathsf{v}_{7}+r^{-z}\sqrt{z+1}\mathsf{v}_{6}-r^{-(z+1)}\mathsf{v}_{5}
    +\frac{2\sqrt{z+1}}{r}\mathsf{v}_{4}\nonumber\\
    +(z-1)\mathsf{v}_{3}-\frac{\mathsf{v}_{2}}{2r^{z+1}}={}\frac{k^{2}\mathsf{v}_{3}}{r^{z+1}}  \label{veq:3}\\
2\sqrt{z+1}(r\mathsf{v}_{8}^{\prime}+(z+2)\mathsf{v}_{8})={}k^{2}\left(\frac{\sqrt{z+1}\mathsf{v}_{6}}{r}-\frac{\mathsf{v}_{2}}{2r^{2}}-\frac{\mathsf{v}_{5}}{r^{2}}-\frac{z\mathsf{v}_{7}}{r}-\mathsf{v}_{7}^{\prime}\right)
   +\omega^{2}\frac{2\sqrt{z+1}\mathsf{v}_{4}}{r^{z}}+k^{2}\omega^{2}\frac{\mathsf{v}_{3}}{r^{2z}}\label{veq:24}\\
   \sqrt{z+1} r\mathsf{v}_{6}^{\prime}-\mathsf{v}_{5}^{\prime}-\frac{\mathsf{v}_{2}^{\prime}}{2}
   -\frac{2\sqrt{z+1}\mathsf{v}_{8}}{r}-(z+1)\mathsf{v}_{7}-\frac{\mathsf{v}_{5}}{r}+\frac{\mathsf{v}_{2}}{2r}
   ={}k^{2}\frac{\mathsf{v}_{7}}{r^{2}}+\omega^{2}\Big(\frac{\mathsf{v}_{3}'}{r^{2z-2}}
   -\frac{\mathsf{v}_{1}}{r}-\frac{\mathsf{v}_{7}}{r^{2z}}\Big)\label{veq:5}\\
 r^{2}\mathsf{v}_{4}^{\prime\prime}+r^{1-z}\mathsf{v}_{8}^{\prime}+r(z+3)\mathsf{v}_{4}^{\prime}
   -(z-2)r^{-z}\mathsf{v}_{8}-2\mathsf{v}_{4}={}k^{2}\Big(\frac{\sqrt{z+1}\mathsf{v}_{1}}{r^{2-z}}
   +\frac{\mathsf{v}_{4}}{r^{2}}+\frac{\mathsf{v}_{6}}{r^{z+1}}+\frac{\sqrt{z+1}\mathsf{v}_{3}}{r^{z}}\Big) \label{veq:6}\\
   r^{2}\mathsf{v}_{1}''-r^{4-2z}\mathsf{v}_{3}''-(z+3)r^{3-2z}\mathsf{v}_{3}'
   -(z-1)\Big(2r^{1-2z}\mathsf{v}_{7}+2r^{1-2z}\sqrt{z+1}\mathsf{v}_{6}+2r^{-z}\sqrt{z+1}\mathsf{v}_{4}\Big)\nonumber\\
   +r(3z+1)\mathsf{v}_{1}'-r^{-2z}\mathsf{v}_{2}=k^{2}\left(\frac{\mathsf{v}_{1}}{r^{2}}-\frac{\mathsf{v}_{3}}{r^{2z}}\right)\label{veq:9}\\
   \sqrt{z+1}(z-1)\Big(r\mathsf{v}_{6}'-\frac{\mathsf{v}_{8}}{r}-\sqrt{z+1}\mathsf{v}_{7}
   +\frac{\mathsf{v}_{5}}{r\sqrt{z+1}}+\frac{\mathsf{v}_{2}}{2r\sqrt{z+1}}\Big)+\frac{\mathsf{v}_{2}'}{2}
   =\nonumber\\
   {}\omega^{2}\left(\frac{(1-z)\mathsf{v}_{1}}{r}-\frac{\mathsf{v}_{1}'}{2}-\frac{\mathsf{v}_{3}'}{2r^{2z-2}}\right)\label{veq:G23}\\
   r^{2}\mathsf{v}_{6}''-\mathsf{v}_{8}'-\sqrt{z+1}r\mathsf{v}_{7}'+(z+3)r\mathsf{v}_{6}'-(z+2)\sqrt{z+1}\mathsf{v}_{7}-\frac{z\mathsf{v}_{8}}{r}
   +\frac{\sqrt{z+1}\mathsf{v}_{5}}{r}+\frac{\sqrt{z+1}\mathsf{v}_{2}}{2r}={}\nonumber\\
   \omega^{2}\left(-\frac{\sqrt{z+1}\mathsf{v}_{1}}{r}
   -\frac{\mathsf{v}_{4}}{r^{z+1}}+\frac{\mathsf{v}_{6}}{r^{2z}}\right)\label{veq:7}\\
   2(z+1)r\mathsf{v}_{8}=k^{2}\left(\mathsf{v}_{6}+r\mathsf{v}_{6}'-2\sqrt{z+1}\mathsf{v}_{7}-\frac{\mathsf{v}_{8}}{r}\right)
   +\omega^{2}\left(r^{1-2z}\mathsf{v}_{8}+r^{2-z}\mathsf{v}_{4}'+r^{1-z}z\mathsf{v}_{4}\right)\label{veq:8}
\end{flalign}
where the corresponding components are $\mathcal{E}_{x_{1}x_{2}}$, $\mathcal{E}_{x_{2}}$, $\mathcal{E}^{2}_{t}$, $(\mathcal{E}^{1}_{y}-\mathcal{E}^{2}_{x})$, 
$\mathcal{E}^{2}_{r}$, $\mathcal{E}^{3}_{t}$, $\mathcal{E}_{tx_{2}}$, $\mathcal{E}_{x_{2}r}$, $\mathcal{E}^{3}_{x_{1}}$, 
$\mathcal{E}^{3}_{r}$. This time there are two constraint equations
\begin{align}
\partial_{r}[\eqref{veq:8}]-\sqrt{z+1}\eqref{veq:24}+\frac{z+1}{r}\eqref{veq:8}+\frac{k^{2}}{r}\eqref{veq:7}-\frac{\omega^{2}}{r^{z}}\eqref{veq:6}&=0,\\
r\partial_{r}[\eqref{veq:G23}]-(z-1)\sqrt{z+1}\eqref{veq:7}-(z-1)\eqref{veq:5}+(z+3)\eqref{veq:G23}-\frac{\eqref{veq:1}}{2r}+\frac{\omega^{2}}{2r}\eqref{veq:9}&=0,
\end{align}
 as the system is now composed of ten equations and eight functions. After we determine $\mathsf{v}_{4}$ algebraically from \eqref{veq:G23}, we expand and solve the system \eqref{veq:1}-\eqref{veq:G23}. 
\endgroup

\end{document}